\documentclass[%
superscriptaddress,
preprint,
 amsmath,amssymb,
 aps,
]{revtex4-2}

\usepackage{titlesec}
\usepackage{graphicx}
\usepackage{dcolumn}
\usepackage{bm}
\usepackage{color}
\usepackage{setspace}
\usepackage{float}
  
\usepackage[colorlinks, linkcolor=blue, anchorcolor=blue, citecolor=magenta]{hyperref}

\def \bU {\mathbf{U}}
\def \Umag {U_{mg}}
\def \bI {\mathbf{I}}
\def \bu {\mathbf{u}}
\def \bx {\mathbf{x}}
\def \bxc {\mathbf{x}_c}
\def \bF {\mathbf{F}}
\def \bT {\mathbf{T}}
\def \be {\mathbf{e}}
\def \br {\mathbf{r}}
\def \bt {\mathbf{t}}

\def \nd {\mathrm{d}}
\def \nexp {\mathrm{exp}}

\def \bOmega {\mathrm{\boldsymbol{\Omega}}}
\def \bsig {\mathrm{\boldsymbol{\sigma}}}
\def \bub {\bar{\bu}}

\def \bn {\mathbf{n}}
\def \cb {\bar{c}}
\def \Ab {\bar{A}}
\def \Mb {\bar{M}}
\def \Db {\bar{D}}
\def \ab {\bar{a}}
\def \bb {\bar{b}}

\def \tb {\bar{t}}

\def \cfb {\bar{c}_f}
\def \etab {\bar{\eta}}

\def \Omegpk {\Omega_{pk}}
\def \Omegmg {\Omega_{mg}}

\def \Vc {\bar{\mathcal{V}}}

\def \dtex {\mathrm{d}}
\def \Pe   {\mathrm{Pe}}
\def \Re   {\mathrm{Re}}
\def \Pecone   {\mathrm{Pe}_c^{(1)}}
\def \Pectwo   {\mathrm{Pe}_c^{(2)}}

\def \lp {\left(}
\def \rp {\right)}
\def \lb {\left[}
\def \rb {\right]}
\def \ln {\left |}
\def \rn {\right |}
\def \lt {\left <}
\def \rt {\right >}

\def \cp  {c^{\prime}}
\def \bup {\bu^{\prime}}

\def \pp  {p^{\prime}}

\def \bsigp {\mathrm{\boldsymbol{\sigma}}^{\prime}}

\def \ch  {\hat{c}}
\def \buh {\hat{\bu}}
\def \bUh {\hat{\bU}}
\def \ph {\hat{p}}
\def \bOmegah {\hat{\mathrm{\boldsymbol{\Omega}}}}
\def \bsigh {\hat{\mathrm{\boldsymbol{\sigma}}}}

\titlespacing*{\subsection}{0pt}{20pt}{10pt}

\begin{document}

\setstretch{1.2} 

\title{Self-propulsion of an elliptic phoretic disk emitting solute uniformly}

\author{Guangpu Zhu}
\affiliation{Department of Mechanical Engineering, National University of Singapore, 117575, Singapore}

\author{Lailai Zhu}
\email{lailai\_zhu@nus.edu.sg}
\affiliation{Department of Mechanical Engineering, National University of Singapore, 117575, Singapore}

\begin{abstract}
Self-propulsion of chemically active droplets and phoretic disks has been widely studied; however, most research overlooks the influence of disk shape on swimming dynamics. Inspired by the experimentally observed prolate composite droplets and elliptical camphor disks, we employ simulations to investigate the phoretic dynamics of an elliptical disk that uniformly emits solutes in the creeping flow regime. By varying the disk's eccentricity $e$ and the P\'eclet number $\Pe$, we distinguish five disk behaviors: stationary, steady, orbiting, periodic, and chaotic. We perform a linear stability analysis (LSA) to predict the onset of instability and the most unstable eigenmode when a stationary disk spontaneously transitions to steady self-propulsion. In addition to the LSA, we use an alternative approach to determine the perturbation growth rate, illustrating the competing roles of advection and diffusion. The steady motion features a transition from a puller-type to a neutral-type swimmer as $\Pe$ increases, which occurs as a bimodal concentration profile at the disk surface shifts to a polarized solute distribution, driven by convective solute transport.
An elliptical disk achieves an orbiting motion through a chiral symmetry-breaking instability, wherein it repeatedly follows a circular path while simultaneously rotating. 
The periodic swinging motion, emerging from a steady motion via a supercritical Hopf bifurcation, is characterized by a wave-like trajectory.
We uncover a transition from normal diffusion to superdiffusion as eccentricity $e$ increases, corresponding to a random-walking circular disk and a ballistically swimming elliptical counterpart, respectively.
\end{abstract}

\maketitle

\section{Introduction} \label{sec:introduc}
Synthetic micro-swimmers have attracted much attention owing to their promising potential in biomedical and bioengineering applications \cite{sitti2015biomedical}, e.g., detection and collection of metal ions \citep{ban2018motion}, targeted controlled drug delivery \citep{kagan2010rapid, tang2020enzyme}, and cancer-cells microsurgery \citep{vyskocil2020cancer}. 
Drawing inspiration from the propulsion strategies of microorganisms in nature, various bio-mimetic swimmers that propel in viscous fluids have been developed \citep{ghosh2009controlled, van2009printed, ahmed2016artificial, soto2021smart}. Unlike their biological counterparts, these synthetic micro-swimmers are commonly powered by external forces or torques coming from the electric, optic, acoustic, or magnetic fields \citep{rao2015force, koleoso2020micro, palagi2016structured}, for example, a sperm-mimicking micro-swimmer with a flexible filament actuated magnetically~\citep{dreyfus2005microscopic}. Despite the rapid development of externally actuated micro-swimmers, some practical difficulties, such as miniaturization and manufacturing of moving parts for certain swimmers, 
have limited their applications in realistic scenarios \citep{ebrahimi2021magnetic, joh2021materials, li2022soft}. 

Unlike externally actuated swimmers, chemically active swimmers convert chemical energy stored internally or extracted from their surroundings into motion \citep{moran2017phoretic}. They can be broadly classified by whether their surface properties, e.g., surface activity and mobility, are anisotropic or isotropic. A classical anisotropic swimmer is the Janus colloid, e.g., the autophoretic Au-Pt Janus colloid \citep{paxton2004catalytic}. The chemically patterned asymmetric colloid typically features two distinct compartments, each composed of a different material or bearing diverse functional groups \citep{lattuada2011synthesis}, which enables asymmetric chemical reactions at the surface. 
The inherent asymmetry allows it to self-generate a concentration gradient, which drives a slip flow inducing net phoretic propulsion, as revealed by experimental \citep{paxton2006chemical, duan2015synthetic, campbell2019experimental}, theoretical \citep{golestanian2007designing, brady2011particle, nasouri2020exact, datt2017active} and numerical \citep{popescu2010phoretic, sharifi2016pair, kohl2021fast} 
studies. The Janus swimmer is typically micro-scale or even smaller, whose phoretic motion is Brownian \citep{michelin2023self}. Its self-propulsion requires a built-in asymmetry in the surface properties. This requirement presents a challenge to the controlled and reproducible manufacturing of Janus colloids, hence hindering their high-throughput production \citep{su2019janus}.

Chemically isotropic swimmers are much easier to manufacture compared to their anisotropic counterparts. A simple and typical representative of such swimmers is a chemically active droplet, e.g., a water droplet slowly dissolving in a surfactant-saturated oil phase, which has been extensively researched since their first experimental realization \citep{izri2014self}. 
These active droplets are generally larger than the Janus microswimmers and have typical radii ranging from 10 to 100 $\mu \mathrm{m}$. Active droplets mainly consist of reacting
droplets \citep{thutupalli2013tuning, kasuo2019start,suematsu2019interfacial} and solubilizing droplets \citep{peddireddy2012solubilization,seemann2016self, hokmabad2021emergence}. 
The former involve chemical reactions producing or changing surfactant molecules at their surface, while the latter feature a micellar dissolution into the surfactant-saturated ambient phase. In both instances, spatial modulation of the surface tension at the droplet interface may potentially induce Marangoni flows. 
Such droplets do not rely on built-in asymmetry like the Janus colloids, but instead attain self-propulsion via an instability spontaneously breaking the spatial symmetry. 
This instability arises from the nonlinear convective transport of solute species by fluid flow, resulting from Marangoni and/or phoretic effects produced by local chemical gradients surrounding the droplet~\citep{morozov2019nonlinear, picella2022confined}. Besides the active droplet, another class of chemically isotropic swimmers is camphor disks that surf at a liquid-air interface \citep{tomlinson1862ii, nakata2006self, suematsu2010mode}. 
In this scenario, camphor molecules dissolved from the disk diffuse into the interface and further the subsurface liquid, and the Marangoni flows resulting from the solute gradient drive the disk to propel~\citep{matsuda2016acceleration, boniface2021role}. 
Notably, the Marangoni flows generated by active droplets are solely at their surface, while those triggered by camphor disks are along the air-liquid interface and depend significantly on the depth of the subsurface liquid~\citep{matsuda2016acceleration, michelin2023self}. 

Active droplets exhibit complex and tunable motion as a result of the nonlinear physico-chemical hydrodynamics \citep{hokmabad2021emergence, li2022swimming}, characterized by the P\'eclet number $\Pe$ as the ratio of flow advection to solute diffusion. 
At low $\Pe$, an isolated droplet remains stationary. \citet{morozov2019self} identified the critical P\'eclet number for an undeformable droplet through a stability analysis, beyond which an unstable dipolar mode of hydrodynamics emerges, driving the spontaneous propulsion of the droplet. 
This critical $\Pe$ is unchanged when the droplet internal flow is neglected, as identified for a chemically isotropic spherical particle initially proposed to mimic an active droplet~\citep{michelin2013spontaneous}.
Besides, the critical $\Pe$ determined for a two-dimensional (2D) undeformable droplet \citep{li2022swimming} is also consistent with that for a phoretic disk \citep{hu2019chaotic}. 
Near the critical $\Pe$, the dipolar mode is the only unstable one \citep{schnitzer2022weakly, peng2022weakly}. However, higher-order modes, e.g., the quadrupolar mode, becomes successively unstable with increasing $\Pe$, leading to the possible coexistence of multiple unstable modes with different polar symmetries \citep{morozov2019nonlinear, hokmabad2021emergence}. 
Accordingly, the active droplet sequentially exhibits quasi-ballistic, unsteady curvilinear, and even chaotic motions \citep{kruger2016curling,suga2018self,hokmabad2021emergence,li2022swimming} as $\Pe$ grows. Analogous behaviors of 2D~\citep{hu2019chaotic, hu2022spontaneous} and three-dimensional (3D) isotropic phoretic particles have also been observed~\citep{chen2021instabilities, hu2022spontaneous, kailasham2022dynamics}. 
Besides an isolated unbounded droplet/particle, the effect of nearby boundaries/fluid interfaces~\citep{thutupalli2018flow, jin2018chemotactic, malgaretti2018self, de2019flow, lippera2020collisions, desai2021instability, dey2022oscillatory, picella2022confined}, 
that of an ambient flow \citep{yariv2017phoretic, dey2021oscillatory, dwivedi2021rheotaxis}, and interaction among multiple droplets/particles \citep{jin2017chemotaxis, lippera2020bouncing, meredith2020predator, nasouri2020exactpho, wentworth2022chemically, hokmabad2022chemotactic, yang2023collective} have been investigated. One specific point we should mention that an active droplet/particle near boundaries \citep{daddi2022diffusiophoretic}, fluid interfaces \citep{malgaretti2018self}, or the other droplet/particle \citep{michelin2015autophoretic} generally exploits geometric asymmetry to propulsion, which is significantly distinct from an isolated droplet/particle.  

Most of the active droplets observed in experiments were weakly deformed and remained spherical, one exception is the very recent work \citep{hokmabad2019topological} reporting the self-propulsion of a prolate composite droplet along its minor axis. That oil droplet trapping two aqueous daughter droplets at the opposing poles of its major axis, suspended in an aqueous surfactant solution. 
The daughter droplets are submicellar, whereas the external aqueous phase is supramicellar. The micellar dissolution at the external oil-water interface induces a self-sustaining surface tension gradient, driving the droplet motion. 
Compared to droplets, solid self-propelling swimmers relying on a similar symmetry-breaking mechanism exhibit greater flexibility in shape. \citet{kitahata2013spontaneous} and \cite{iida2014theoretical} 
 experimentally and theoretically investigated the spontaneous motion of an elliptical camphor disk at the air-liquid interface, and found that the disk swam along its minor axis resembling the swimming prolate droplet \citep{hokmabad2019topological}. 
 \citet{shimokawa2022hula} observed that an elliptical camphor-coated paper disk exhibited spontaneous rotation at a constant angular velocity. 
Motivated by these non-spherical phoretic swimmers with uniform chemical reactions at their surface, especially \citet{kitahata2013spontaneous} and \citet{hokmabad2019topological}, here, we theoretically and numerically explore, in the creeping flow regime, the instability-driven spontaneous propulsion of an elliptical phoretic disk that releases chemical species uniformly. 
We perform a linear stability analysis (LSA) to investigate the onset of instability, and direct numerical simulations to explore the swimming behavior of the phoretic disk.

This paper is organized as follows. We describe the problem setup, assumptions, and governing equations in Section \ref{sec:prob}. The implementation for the LSA is introduced in Section \ref{sec:LSA}, followed by Section \ref{sec:results} demonstrating numerical and theoretical results. Finally, we conclude our observations and provide some discussion in Section \ref{sec:conclu}.

\section{Problem setup, governing equations and methodology} \label{sec:prob}
\subsection{Problem setup and governing equations}
\begin{figure}[H]
\centering
\includegraphics[scale=0.85]{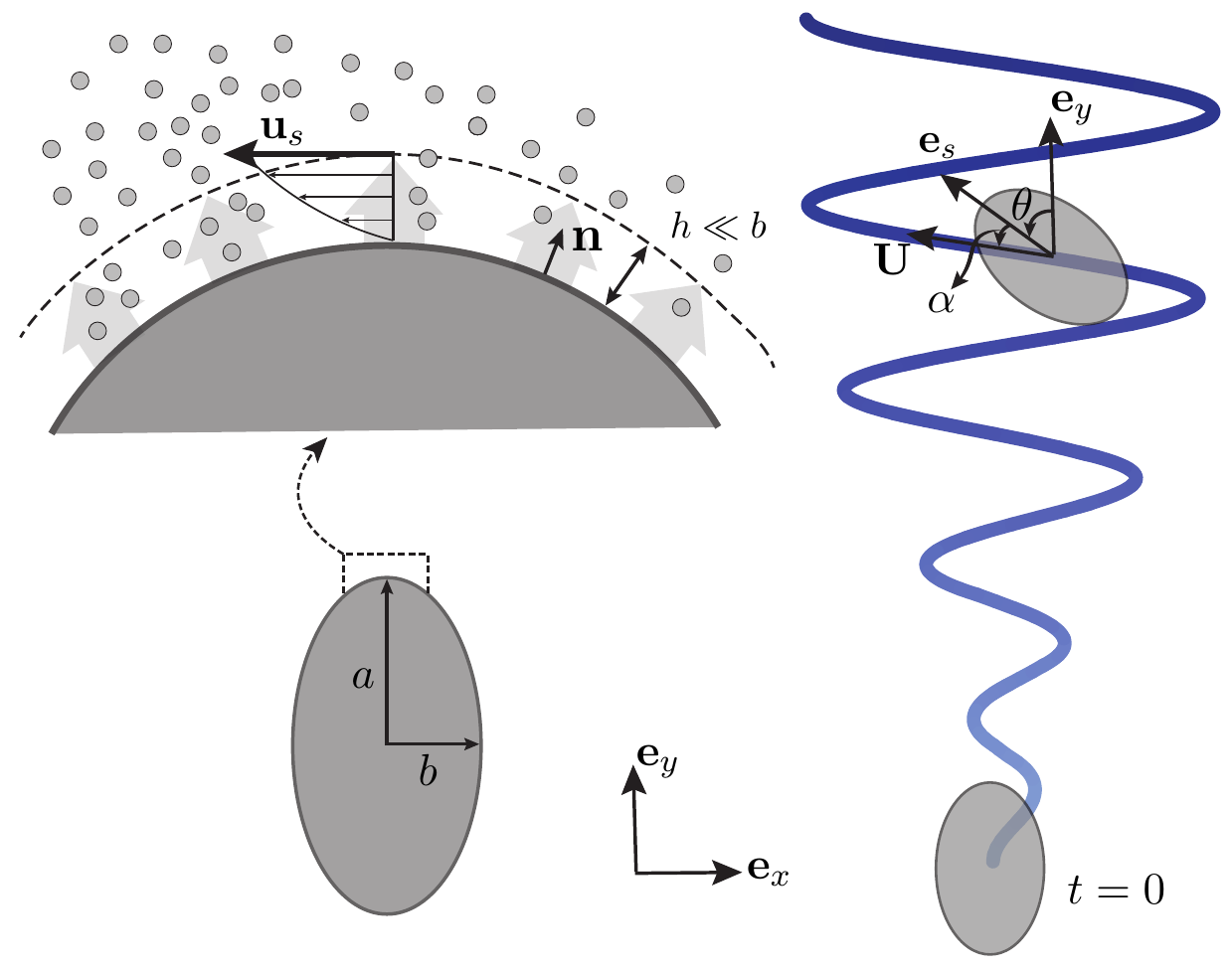}
\caption{Self-propulsion of an elliptical disk uniformly releasing chemical solutes in a Newtonian solvent. $\be_{xy}$ denotes the laboratory frame. 
The disk moves along an undulatory path with a translational velocity of $\bU$ and the color of the path is coded by the time $t$.
The disk orientation $\be_s$ coinciding with its major axis deviates from $\be_{y}$ and $\bU$ by angles $\theta$ and $\alpha$, respectively. 
The inset shows the induced slip velocity $\bu_s$ by local solute gradients within a thin boundary layer of thickness $h \ll b$. $\bn$ denotes the unit normal vector pointing away from the disk surface. All the variables here are dimensionless.}
\label{fig:sketch}
\end{figure}

We consider a chemically active elliptical disk emitting or absorbing solute molecules uniformly in an incompressible Newtonian fluid of dynamic viscosity $\etab$ (see figure \ref{fig:sketch}). 
From here on, the bar indicates dimensional variables unless otherwise mentioned. 
The semi-major and semi-minor axes of the disk are $\ab$ and  $\bb\leq \ab$, respectively, and $\cfb = \sqrt{\ab^2-\bb^2}$ denotes the half of its focal length. Hence, the disk shape can be characterized by the
eccentricity $e = \cfb/\ab$, which amounts to $0$ or approaches $1$ as the disk becomes circular or needle-like. 
We choose the major axis of the disk to denote its orientation $\be_s = \sin \theta \be_{x}+\cos{\theta} \be_{y} $, which is characterized by its angular deviation from $\be_{y}$. 

We now describe the phoretic dynamics of the elliptical disk and the associated governing equations \citep{anderson1989colloid, michelin2013spontaneous}. The disk's surface $\Gamma_d$ 
uniformly emits or absorbs solutes with a constant rate $\Ab$ (activity), hence, 
\begin{equation}
   \left. \Db\bn \cdot \nabla \cb \right |_{\Gamma_d}= -\Ab.
\label{flux_c}
\end{equation}
Here, $\cb$ is the solute concentration, $\Db$ denotes the molecular diffusivity of the solute, and $\bn$ is the unit outward normal at the surface. Positive or negative $\Ab$ corresponds to the solute emission or absorption at the disk surface, respectively. The solute interacts with the disk surface through a short-range potential and here we focus on the classical thin-interaction-layer limit $h \ll \bb$, with $h$ the thickness of the interaction layer. Within this layer, the slip velocity along the disk surface induced by the local tangential solute gradients \citep{anderson1989colloid} reads
\begin{equation}
   \left. \bub_s \right |_{\Gamma_d} = \Mb \nabla_s \cb,
\end{equation}
where $\bub_s$ indicates the slip velocity, $\nabla_s = \lp \bI -\bn \bn \rp \cdot \nabla$ is the surface gradient operator, and $\Mb$ denotes the phoretic mobility coefficient. 
The coefficient $\Mb$ determines the chemotactic direction of the phoretic swimmer. For active droplets, $\Mb$ can be adjusted by tuning the relationship between surface tension and chemical solute, as shown by recent experiments~\citep{wentworth2022chemically}: $\Mb>0$ when they are positively correlated and 
$\Mb<0$ when negatively correlated. Prior studies~\citep{michelin2013spontaneous, hu2019chaotic} revealed that spontaneous symmetry-breaking propulsion is present only when $\Ab\Mb>0$; in other cases, the disk remains stable. Without loss of generality, our analysis focuses on 
$\Ab>0$ and $\Mb>0$, but the results will remain in the converse scenarios.
Also, we assume that the inertia of both fluid and disk is negligible compared to the viscous force because the Reynolds number $\Re$ is typically small in the experiments \citep{peddireddy2012solubilization, maass2016swimming, de2019flow, hokmabad2021emergence, michelin2023self, hokmabad2022chemotactic}. Hence, the fluid flow surrounding the disk can be described approximately by the Stokes equation. 

In the following, we introduce the dimensionless governing equations. 
 We choose $\Ab \Mb/\Db$, $\bb$, and $\bb\Ab/\Db$, respectively, as the characteristic velocity $\Vc$, length, and concentration
for the nondimensionalization. 
All variables below are dimensionless unless otherwise specified. The dimensionless equations for the velocity $\bu$, pressure $p$, and concentration $c$ are 
\begin{equation}
    \nabla \cdot \bsig = \mathbf{0}, \;\;\; \nabla \cdot \bu = 0,
\label{NS}
\end{equation}
\begin{equation}
    \frac{\partial c}{\partial t} + \bu \cdot \nabla c = \frac{1}{\Pe} \Delta c,
\label{adv_diff}
\end{equation}
where the P\'eclet number $\Pe = \Ab\Mb\bb/\Db^2$ measures the ratio of flow advection to solute diffusion and $\bsig = -p\bI+  \nabla \bu + \lp\nabla \bu\rp^{T}  $ is the hydrodynamic stress tensor. At the disk surface, the constant flux boundary condition \eqref{flux_c} for the concentration reads
\begin{equation}
    \left. \bn \cdot \nabla c \right |_{\Gamma_d} = -1.
\label{c_par}
\end{equation}
It is known that equation (\ref{adv_diff}) does not support a steady-state solution within a 2D infinite domain due to the logarithmic divergence \citep{sondak2016can, yariv2017two, kailasham2023non}. To resolve this issue, we consider a circular fluid domain with a finite radius $R$ and prescribe on its exterior $\Gamma_o$
\begin{equation}
    \left. c \right |_{\Gamma_o} = 0.
\label{c_far}
\end{equation}
Following \citet{hu2019chaotic} and \citet{li2022swimming}, here we set $R = 200$. In fact, we also examine the convergence of numerical results with respect to the domain size $R$ and reveal that the chosen $R$ is sufficiently large to ensure that the disk motion remains unaffected, as depicted in figure \ref{fig:Rdep} of Appendix C.

We perform numerical simulations in the frame co-moving with the disk center. Hence, the boundary conditions for the velocity at the disk surface $\Gamma_d$ and the outer boundary $\Gamma_o$ are
\begin{subequations}
\begin{align}
    \bu  |_{\Gamma_d} & = \bu_s + \bOmega \times \lp \bx_s - \bxc \rp, \\
    \bu  |_{\Gamma_o} & = -\bU,
\end{align}
\label{u_boud}
\end{subequations}
where $\bu_s = \nabla_s c$ is the slip velocity at the disk surface, $\bx_s$ and $\bx_c$ denote the coordinates of a general point at the disk surface and the disk center, respectively. Here, $\bU$ and $\bOmega$ denote the translational and rotational velocities of the disk, respectively, which are determined by the force-free and torque-free conditions \citep{lauga2009hydrodynamics}
\begin{subequations}
\begin{align}
    \bF  = \int_{\Gamma_d} \bn \cdot \bsig \;\dtex l & = \mathbf{0}, \\
    \bT  = \int_{\Gamma_d} \lp \bx_s - \bxc \rp \times \lp \bn \cdot \bsig \rp \dtex l & = \mathbf{0}.    
\end{align}
\label{force_free}
\end{subequations}
Namely, the total force and torque exerted on the disk swimmer are zero. The initial condition is zero velocity, pressure, and concentration within the domain. 
Equations (\ref{NS})-(\ref{force_free}) form the complete set of governing equations for an elliptical phoretic disk in the frame of co-moving with the disk.

\subsection{Numerical method}
We numerically solve the governing equations, using a finite-element-method solver implemented in the commercial package COMSOL Multiphysics (I-Math, Singapore). We adopt the moving mesh technique to tackle the deformation of the fluid domain caused by disk rotation. 
Taylor-Hood and quadratic Lagrange elements are employed to discretize the flow field, $\lp \bu, p\rp$, and the concentration $c$, respectively. The computational domain is discretized by approximately $85000 - 127000$ triangular elements, and the mesh is locally refined near the disk. Our COMSOL implementations are extensively validated against several published datasets, as shown in Appendix \ref{sec:appenA}.

\section{Linear stability analysis} \label{sec:LSA}
Prior studies reveal that a chemically isotropic disk/droplet possesses a stable stationary state at a sufficiently low $\Pe$.
As $\Pe$ grows beyond a critical value, an instability arises, leading the disk/droplet to swim autonomously \citep{hu2019chaotic, morozov2019self, li2022swimming}. 
Our numerical results indicate that the elliptical disk here exhibits analogous behavior, hence we conduct a LSA to examine the onset of instability at a critical P\'eclet number $\Pecone$. 

We first decompose the space- ($\bx$) and time-dependent state variables $\lp c, \bu, p \rp$ into the sum of a base state and a perturbation state as
\begin{subequations}
\begin{align}
  c \lp \bx, t \rp &= c_b \lp \bx \rp + \cp \lp \bx, t \rp, \\
 \bu \lp \bx, t \rp &= \bu_b \lp \bx \rp + \bup \lp \bx, t \rp, \\
 p \lp \bx, t \rp & = p_b \lp \bx \rp + \pp \lp \bx, t \rp, 
\end{align}
\label{base_per}
\end{subequations}
where the subscript $b$ denotes base-state fields and the primed variables are infinitesimal perturbations. The base state can be obtained numerically.
By substituting \eqref{base_per} into \eqref{NS} and \eqref{adv_diff}, and retaining linear terms, we obtain
\begin{equation}
    \nabla \cdot \bsigp  = \mathbf{0}, \;\; \nabla \cdot \bup = 0,
\label{adv_diff_p}
\end{equation}
\begin{equation}
     \frac{\partial \cp}{\partial t} + \bu_b \cdot \nabla \cp + \bup \cdot \nabla c_b = \frac{1}{\Pe} \Delta \cp.
\label{NS_p}
\end{equation}
Note that $\bu_b \lp \bx \rp = \mathbf{0}$ for a stationary circular disk, while for its elliptical counterpart, $\bu_b \lp \bx \rp \neq \mathbf{0}$ due to the anisotropic concentration distribution at the disk surface (see figure \ref{fig:phase_concen}a).
The perturbations are assumed to vary exponentially in time with a complex growth rate $\lambda = \lambda_r + i\lambda_i$, i.e., 
\begin{subequations}
\begin{align}
  \cp \lp \bx, t \rp &= \ch \lp \bx \rp \nexp \lp \lambda t \rp  , \\
 \bup \lp \bx, t \rp &= \buh \lp \bx \rp \nexp \lp \lambda t \rp, \\
 \pp \lp \bx, t \rp & = \ph \lp \bx \rp \nexp \lp \lambda t \rp. 
\end{align}
\label{growth_rate}
\end{subequations}
Consequently, equations \eqref{adv_diff_p} and \eqref{NS_p} can be reformulated to
\begin{equation}
    \nabla \cdot \bsigh  = \mathbf{0}, \;\; \nabla \cdot \buh = 0,
\label{NS_h}
\end{equation}
\begin{equation}
     \lambda \ch + \bu_b \cdot \nabla \ch + \buh \cdot \nabla c_b = \frac{1}{\Pe} \Delta \ch.
\label{adv_diff_h}
\end{equation}
Substituting them with \eqref{base_per} into \eqref{c_par} to \eqref{u_boud} enables us to derive the boundary conditions for $\ch$ and $\buh$ at the disk surface and outer boundary 
\begin{equation}
     \left. \bn \cdot \nabla \ch \right |_{\Gamma_p} = 0, \;\;\; \left. \ch \right |_{\Gamma_o} = 0,
\label{c_boud_h}
\end{equation}
\begin{equation}
     \left. \buh \right |_{\Gamma_d} = \nabla_s \ch + \bOmegah \times (\bx_s-\bx_c), \;\;\; \left. \buh \right |_{\Gamma_o} = -\bUh.
\label{u_boud_h}
\end{equation}
Note that $\bU_b$ and $\bOmega_b$ disappear in \eqref{u_boud_h} corresponding to a stationary disk of the base state. The force-free and torque-free conditions still hold as
\begin{subequations}
\begin{align}
    \hat{\bF}  & = \int_{\Gamma_d} \bn \cdot \bsigh \;\dtex l = \mathbf{0}, \\
    \hat{\bT}  & = \int_{\Gamma_d} \lp \bx_s - \bxc \rp \times \lp \bn \cdot \bsigh \rp \dtex l = \mathbf{0}.    
\end{align}
\label{force_torque_h}
\end{subequations}
Equations \eqref{NS_h}-\eqref{force_torque_h} define an eigenvalue problem. The stability of the base state is determined by the eigenvalue with the largest real part $\lambda_r^0$, namely the leading eigenvalue, and the corresponding perturbations $\lp \buh, \ph, \ch \rp^T$ are called leading eigenmodes. The base state is stable when $\lambda_r^0 < 0$ but unstable when $\lambda_r^0 > 0$. The P\'eclet number at which $\lambda_r^0 = 0$ is precisely the critical P\'eclet number $\Pecone$ signifying the transition from a stationary state to steady propulsion. Unless stated otherwise, the eigenvalues mentioned below refer to the leading eigenvalues, and the superscript $0$ is omitted for simplicity. We solve the eigenvalue problem with COMSOL using the eigenvalue solver ARPACK. The validation of our approach is demonstrated in Appendix \ref{sec:appenA}.

\section{Results} \label{sec:results}

\subsection{Diverse behaviours of an elliptical phoretic disk}
\begin{figure}[H]
\centering
\includegraphics[scale=0.55]{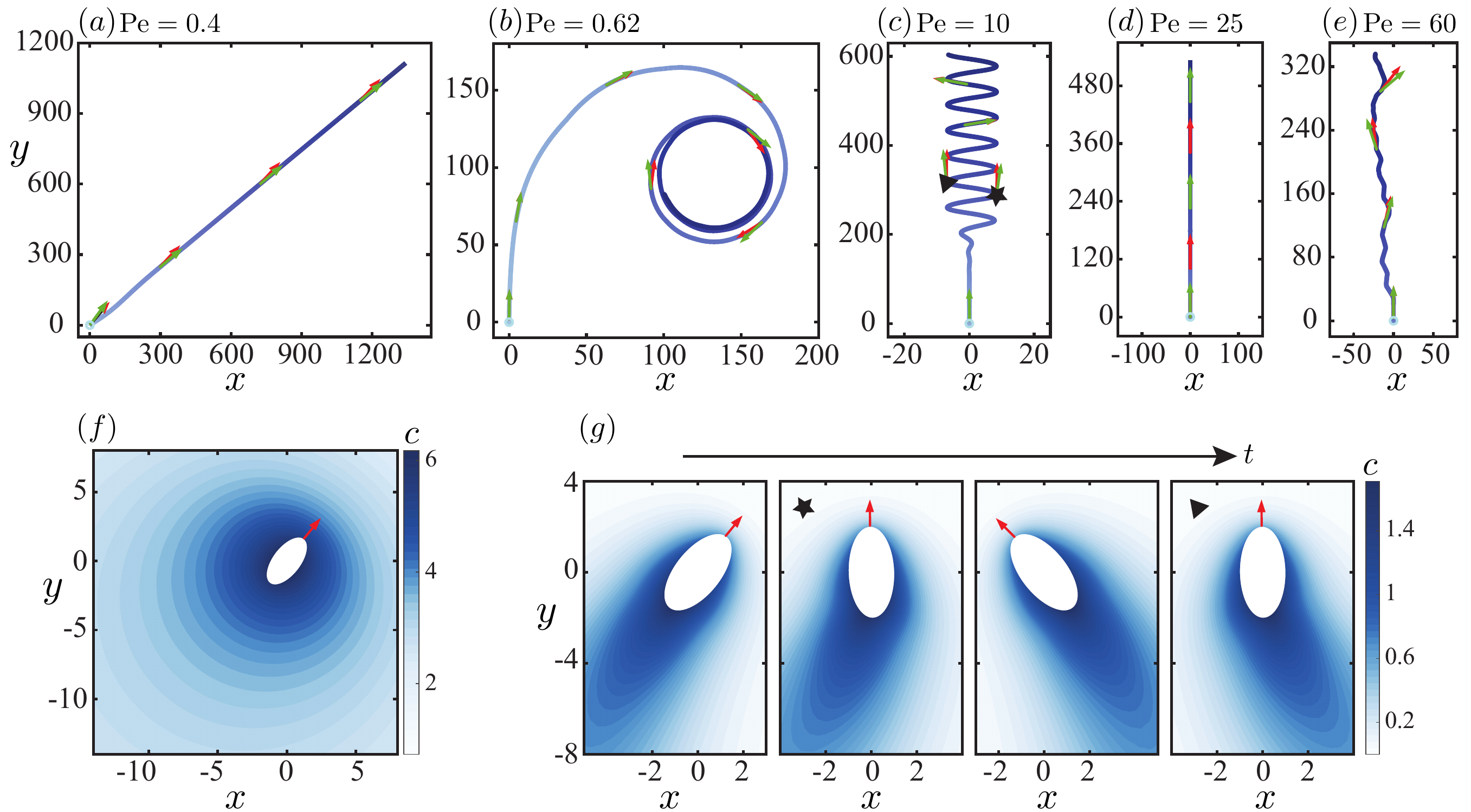}
\caption{An elliptical disk of eccentricity $e = 0.87$ follows typical trajectories depending on $\Pe$:  (a) steady, (b) orbiting, (c) periodic, (d) steady, and (e) chaotic. The color of trajectories is coded by the time $t$. Green and red arrows denote the direction of the translational velocity $\bU$ and the disk orientation $\be_s$, respectively. (f) Polarized concentration distribution with respect to $\be_s$ at $\Pe = 0.4$. (g) Periodic pendulum-like swinging of the disk at $\Pe =10$ in the frame co-moving with its center. The star and triangle symbols marked in (g) denote the two distinct moments when the disk reaches the peak and trough of its trajectory, respectively, as shown in (c).
}
\label{fig:trajec}
\end{figure}

By increasing P\'eclet number $\Pe$, we demonstrate the diverse $\Pe$-dependent behaviors of a disk with the eccentricity $e=0.87$. 
For a sufficiently small $\Pe$ below a critical value $\Pecone$, the disk undergoes transient rotation before recovering a stationary state.
When $\Pe$ goes above $\Pecone$, e.g., $\Pe = 0.4$, the stationary state transits into the steady propulsion, in which the fore-aft symmetry in the concentration profile is broken (see figure \ref{fig:trajec}f),
and the resultant concentration polarity induces a rectilinear motion with a constant swimming speed, as depicted in figure \ref{fig:trajec}(a). 
Further increasing $\Pe$ to $0.62$, the directed motion loses its stability, leading to a secondary instability characterized by spontaneous chiral symmetry breaking. Consequently, the disk repeatedly traces a circular path, exhibiting spontaneous rotation (clockwise, in this instance) as depicted in figure \ref{fig:trajec}(b).
 This regime featuring circular trajectory and self-rotation is called the orbiting regime. 
\citet{hu2019chaotic} and \citet{li2022swimming} observed analogous circular trajectories executed by circular swimmers without self-rotation.
At $\Pe = 10$, the elliptical disk favors a wave-like trajectory after a transient period of swimming straightforward (see figure \ref{fig:trajec}c). From the viewpoint of the frame co-moving with the disk center, the disk periodically swings like a pendulum (refer to figure \ref{fig:trajec}g).
This swimming behavior is analogous to that exhibited by an active prolate double-core droplet, which moves along an undulatory trajectory \citep{hokmabad2019topological}. 
\citet{li2022swimming} also reported that a high-$\Pe$ active drop swims along a periodic zigzag trajectory.
As $\Pe$ increases to $25$, the elliptical disk surprisingly recovers to steady propulsion, as illustrated in figure \ref{fig:trajec}(d). In correspondence, chiral symmetry recovers. 
Unlike the ballistic motion after rotation at low $\Pe$, 
the disk here travels straight from the onset of instability without any rotation. 
The steady propulsion becomes unstable at a higher $\Pe$, e.g., $\Pe = 33$, accordingly, the disk swims straightforwardly at an oscillating speed (see figure \ref{fig:pe33} in Appendix B). \citet{hu2022spontaneous} and \citet{kailasham2022dynamics} identified the similar motion of a 3D isotropic phoretic particle in an axisymmetric setup, where the particle does not rotate, and the flow and concentration fields are symmetric about an axis parallel to the particle's translational direction.
Figure \ref{fig:trajec}(e) indicates that the disk
enters into a chaotic regime characterized by an erratic trajectory at $\Pe = 60$. In contrast to frequent intermittency and random walk occurring in the chaotic regime for a circular disk, as observed by \citet{hu2019chaotic}, the change of velocity in both direction and magnitude experienced by the elliptical disk is not drastic, yielding a less chaotic trajectory. 

\begin{figure}[H]
\centering
\includegraphics[scale=0.75]{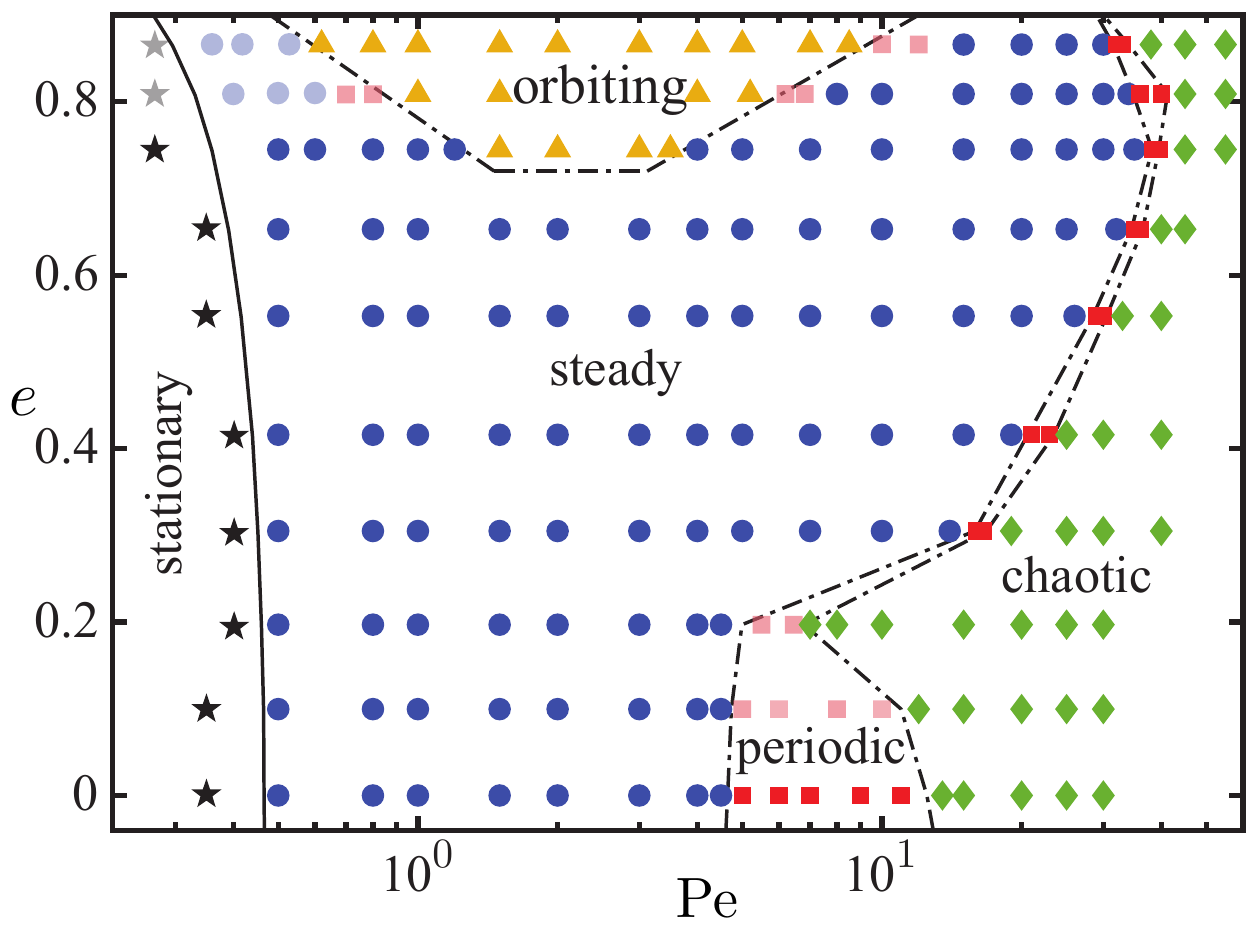}
\caption{Phase diagram characterizing the behaviors of an elliptical phoretic disk depending on its eccentricity $e$ and the P\'eclet number $\Pe$. It shows five regimes: stationary, steady, orbiting, periodic, and chaotic. The solid line denotes the LSA prediction.
}
\label{fig:phase_diag}
\end{figure}

Having observed distinct swimming behaviours of a disk with a specific eccentricity at varying $\Pe$, we then systematically examine how the eccentricity $e$ affects these $\Pe$-dependent behaviours.
The phase diagram in figure \ref{fig:phase_diag} shows that shifts in $\Pe$-dependent behaviours of an elliptical disk with  $e$ below $0.75$, resemble those of a circular counterpart ($e = 0$). The elliptical disk successively executes stationary, steady, periodic, and chaotic motion with increasing $\Pe$. As $e$ exceeds $0.75$, the disk exhibits richer dynamics: the orbiting motion emerges with broken chiral symmetry. In fact, we have also explored the scenarios at $e > 0.9$, e.g., $e = 0.96$, and observed that the swimming dynamics of the elliptical disk are almost dominated by orbiting and chaotic regimes (not shown here). These two regimes have been revealed in the current phase diagram, hence the characteristic locomotory modes can be well captured in the range of $e$ considered. 

We further discern from the phase diagram that the variation of $e$ categorizes certain identified swimming patterns into two types. In the stationary or steady swimming regime, the aforementioned observations indicate that the disk experiences transient rotation before recovering a stationary state or retaining a steady motion (see figure \ref{fig:trajec}). Nevertheless, these situations are only applicable at a higher $e$, clearly distinguished by light-colored symbols. When $e < 0.75$, the disk remains stationary or swims straight without any rotation. Besides, two types of periodic motions are depicted in the phase diagram: 1) a disk swings along a wavy trajectory (light-colored squares); 
2) a disk swims straight at an oscillating swimming speed (dark-colored squares). The former and the latter are termed swinging and straight periodic motions, respectively. 

 \subsection{Spontaneous steady propulsion triggered by instability} 

\begin{figure}[H]
\centering
\includegraphics[scale=0.6]{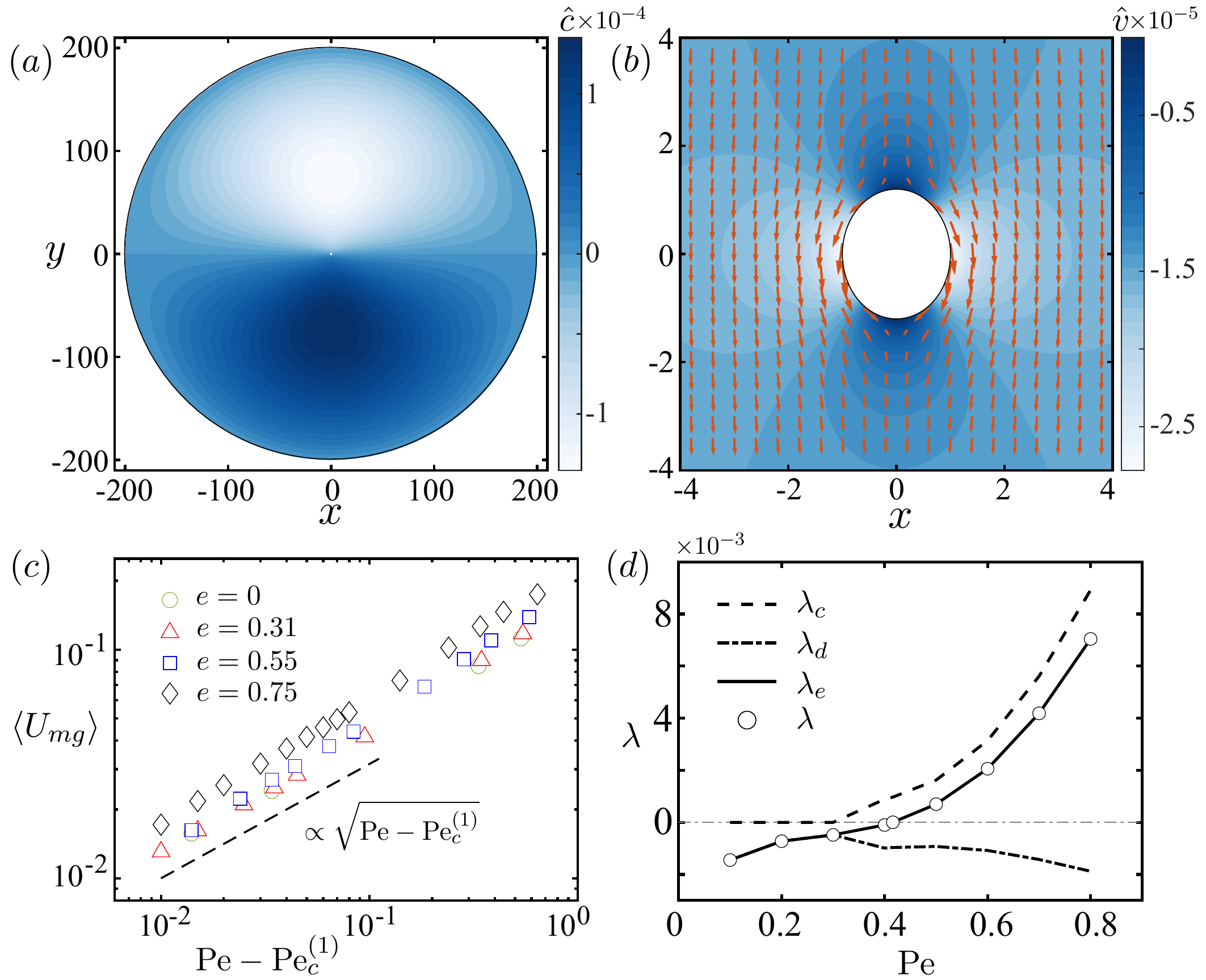}
\caption{Eigenmodes at the critical P\'eclet number $\Pecone \approx 0.42$ for an elliptical disk with $e = 0.55$. The eigenmode is characterized by (a) the perturbation concentration $\ch$, and (b) the perturbation velocity $\buh$ (red arrows) and its $y$ component $\hat{v}$ (color map). (c) Dependence of time-averaged disk speed $\lt \Umag \rt$ on $\Pe-\Pecone$ at varying $e$. In the vicinity of $\Pecone$, $\lt \Umag \rt$ is proportional to $\sqrt{\Pe-\Pecone}$. (d) $\Pe$-dependent growth rate $\lambda$ based on the LSA and that $\lambda_e= \lambda_c + \lambda_d$ derived from the concentration perturbation equation (\ref{adv_diff_h}). The latter comprises the contributions $\lambda_c$ and $\lambda_d$ from advection and diffusion, respectively.} 
\label{fig:eigen}
\end{figure} 

Upon gaining a general understanding of the phase diagram, we then explore the detailed swimming dynamics within each regime. We first focus on the transition from the stationary to steady swimming regime. 
The disk is stationary in the base state. 
As $\Pe$ exceeds $\Pecone$, an instability arises, and the disk sets into steady propulsion along its major axis. Here, we perform a LSA, as introduced in Section \ref{sec:LSA}, to quantitatively identify $\Pecone$ and its dependence on the disk shape. $\Pecone$ predicted by the LSA monotonically decreases with $e$, as depicted by the solid line in the above phase diagram. 
Figure \ref{fig:eigen}(a) depicts the concentration field $\ch$ of the eigenmode at $\Pecone \approx 0.42 $ for $e = 0.55$. 
We see that the symmetry of the base concentration field $\ch$ is broken in the direction of the major axis at $\Pecone$. 
The fore-aft asymmetric concentration distribution induces a downward slip flow, as shown by the flow field $\buh$ of the eigenmode in figure \ref{fig:eigen}(b), driving the steady motion of the disk along its major axis. 
In close proximity to $\Pecone$, the swimming speed $\Umag$ is proportional to $\sqrt{\Pe-\Pecone}$ (see figure \ref{fig:eigen}c), implying that the instability occurs through a supercritical pitchfork bifurcation. This is parallel to the observation of \citet{hu2019chaotic} and \citet{li2022swimming}. 

We further probe the physical mechanism underlying the instability by adopting an approach resembling the energy budget analysis \citep{bendiksen1985motion, abubakar2022linear}. By taking an inner product denoted by $\lp \cdot, \cdot \rp$ of the equation (\ref{adv_diff_h}) with $\ch$ in the $L^2$ space, we arrive
\begin{equation}
     \lp \lambda \ch, \ch \rp + \lp \bu_b \cdot \nabla \ch, \ch \rp + \lp \buh \cdot \nabla c_b, \ch \rp = \frac{1}{\Pe} \lp \Delta \ch, \ch \rp.
\label{adv_diff_inner}
\end{equation}
Using integration by parts, the divergence-free condition in (\ref{NS_h}), and the boundary condition (\ref{c_boud_h}) and (\ref{u_boud_h}), we derive
\begin{subequations}
\begin{align}
    \lp \bu_b  \cdot \nabla \ch, \ch \rp  & = \frac{1}{2} \lp \bn \cdot \bu_b, \ch^2 \rp_{\Gamma_p}  + \frac{1}{2} \lp \bn \cdot \bu_b, \ch^2 \rp_{\Gamma_o}  = 0, \\
     \lp \buh \cdot \nabla c_b, \ch \rp & = -\lp c_b, \buh \cdot \nabla \ch \rp, \\ \frac{1}{\Pe} \lp \Delta \ch, \ch \rp & = -\frac{1}{\Pe} \parallel \nabla \ch \parallel ^2,
\end{align}
\label{inter_parts}
\end{subequations}
with $\parallel \cdot \parallel$ denoting the $L^2$-norm. By substituting (\ref{inter_parts}) into (\ref{adv_diff_inner}), we obtain
\begin{equation}
     \lambda_e  = \lambda_c+\lambda_d,
\label{eigen_h}
\end{equation}
with 
\begin{subequations}
\begin{align}
     \lambda_c & = \frac{\lp c_b, \buh \cdot \nabla \ch \rp}{\lp 1, \ch^2 \rp}, \\
     \lambda_d & =  - \frac{\parallel \nabla \ch \parallel^2}{\lp \Pe, \ch^2 \rp}.
\label{con_convec_diff}
\end{align}    
\end{subequations}
Here, the growth rate $\lambda_e$ is introduced in (\ref{eigen_h}) to be distinguished from $\lambda$ obtained by the LSA. $\lambda_c$ and $\lambda_d$ represent the contributions of advection and diffusion to $\lambda_e$, respectively. For a steady motion, the imaginary part of the growth rate vanishes, thus, the growth rate only has its real part, e.g., $\lambda = \lambda_r$. Figure \ref{fig:eigen}(d) shows that $\lambda_e$ and $\lambda$ lie on top of each other, giving us the confidence to analyze the dominant physical ingredient that drives the instability using (\ref{eigen_h}). 
We naturally infer from (\ref{eigen_h}) that $\lambda_c$ is responsible for $\lambda_e$ turning positive by realizing that $\lambda_d$ is consistently negative, as confirmed by figure \ref{fig:eigen}(d).
Hence, as anticipated, advection drives the instability and diffusion dampens the perturbation, the balance between them dominates the phoretic dynamics of the 
 elliptical disk: at small $\Pe$, diffusion dominates and its homogenizing effect maintains a fore-aft symmetric solute distribution. As $\Pe$ grows beyond $\Pecone$, advection suppresses diffusion and amplifies the asymmetric solute disturbance. The slip flows triggered by the asymmetric concentration distribution drive the disk to swim spontaneously. 

\begin{figure}[H]
\centering
\includegraphics[scale=0.5]{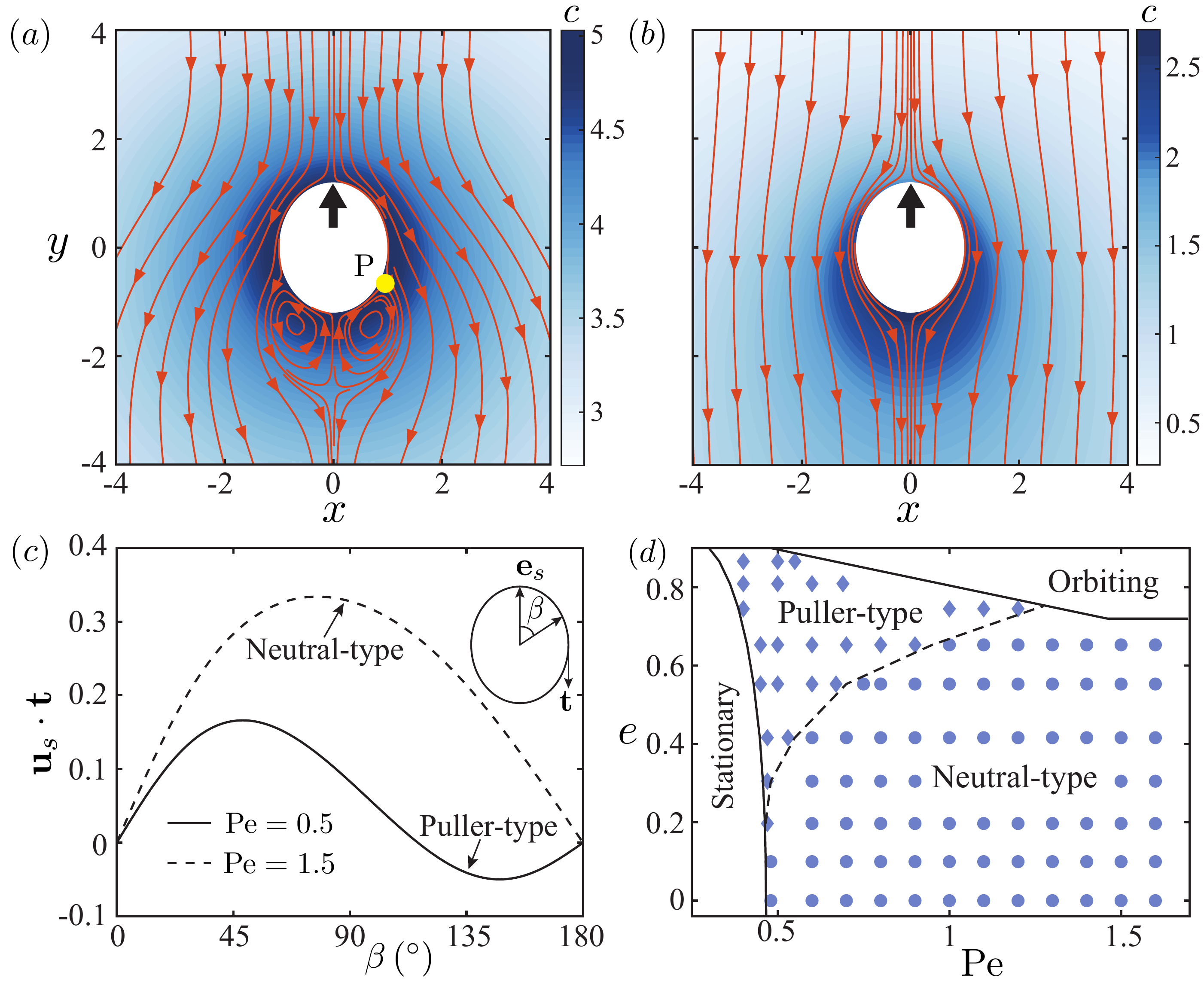}
\caption{(a) A puller-type steady swimmer at $\Pe = 0.5$ transitions to (b) a neutral-type counterpart at $\Pe = 1.5$, where the eccentricity $e$ is $ 0.55$. The flow fields are shown in the frame co-moving with the disk. Black arrows denote its swimming direction.
The color maps indicate the distribution of $c$.  The stagnation point $\mathrm{P}$ coincides with the peak of $c$ at the disk surface. (c) Polar velocity magnitude $\bu_s \cdot \bt$ at the disk surface, following the definition in \citet{downton2009simulation}. Here, $\beta$ is the polar angle with respect to the disk orientation $\be_s$, with $\bt$ the corresponding unit tangent vector.
(d) $\Pe-e$ phase diagram shows puller-type and neutral-type steady swimmers demarcated by the dashed line. Solid lines separate regimes identified in figure \ref{fig:phase_diag}. 
}
\label{fig:phase_puller}
\end{figure} 

Next, we analyze the steady propulsion of the autophoretic disk. We show in figure \ref{fig:phase_puller}(a) and (c) that the elliptical disk of $e = 0.55$ swims as a puller attracting the fluid from its front and rear at $\Pe = 0.5$. 
In contrast, it becomes a neutral-type swimmer at $\Pe=1.5$, as indicated in figure \ref{fig:phase_puller}(b) and (c). 
In retrospect, \citet{suda2021straight} reported an analogous transition from 
a puller swimming straight into a pusher-type swimmer executing unsteady curvilinear motion when $\Pe$ increases. 
In that case, the droplet motion is triggered by the concentration gradient caused by a point source of surfactant at the surface. Besides, \citet{li2022swimming} observes that an active drop transits from a steady pusher to a mixed pusher-puller propelling unsteadily as $\Pe$ grows. 
In the case of these droplet swimmers, the switching of their disturbance flow occurs as they go from steady to unsteady motion. 
\citet{morozov2019nonlinear} describes a $\Pe$-dependent neutral-to-pusher transition for a 3D axisymmetric droplet impelling steadily. Notably, this transition occurs without a shift in the droplet's swimming pattern, resembling that exhibited by the disk swimmer here. We further examine the flow field in the frame co-moving with the disk throughout the whole steady swimming regime and summarize these results in a phase diagram in figure \ref{fig:phase_puller}(d).
It is found that a circular ($e = 0$) or nearly-circular disk can only be a neutral swimmer, and the puller-to-neutral transition emerges as $e\geq 0.20$. The critical P\'eclet number $\Pe_t$ signifying this transition depends monotonically on $e$, as indicated by the dashed line. When $e > 0.65$, the transition disappears and the disk solely swims as a puller. In fact, this peculiar transformation can be understood by examining the solute distribution at the disk surface, as discussed below. 
For visualization purpose, we normalize the concentration via $\Tilde{c} = 0.3 \lp c_{max}-c \rp/\lp c_{max}-c_{min} \rp$, where $c_{max}$ and $c_{min}$ denote the maximum and minimum concentrations at the disk surface, respectively.

\begin{figure}[H]
\centering
\includegraphics[scale=0.56]{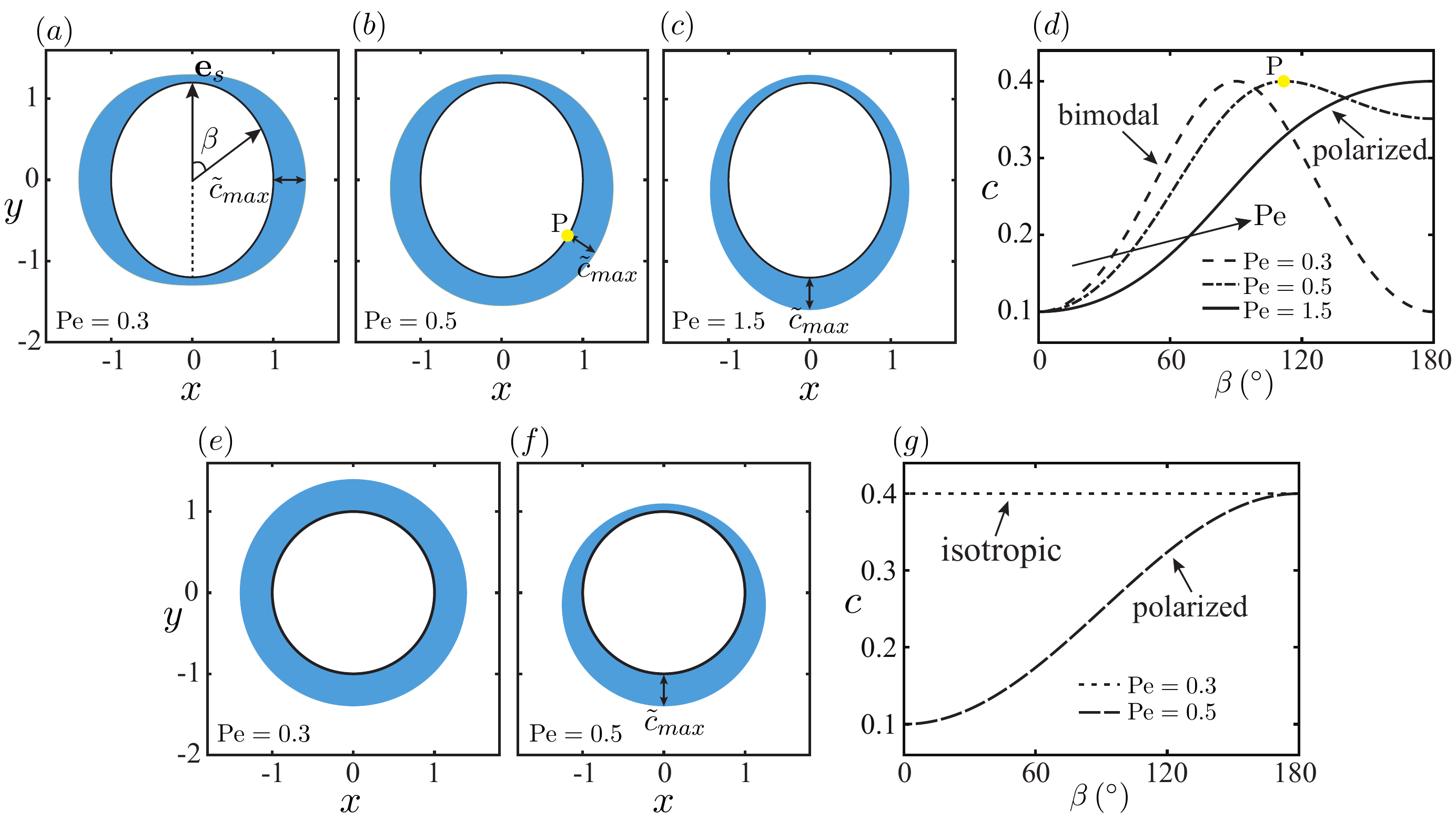}
\caption{Normalized concentration at the surface of a disk swimming steadily: 
(a)-(c) for an elliptical disk with $e = 0.55$ at $\Pe = 0.3$, $0.5$, and $1.5$, respectively; (e)-(f) for a circular disk at $\Pe = 0.3$ and $0.5$, respectively. $c_{max}$ denotes the peak concentration at the disk surface. The stagnation point $\mathrm{P}$ coincides with the location of $\Tilde{c}_{max}$ at $\Pe = 0.5$. (d): transition from a bimodal to polarized concentration distribution at the surface of an elliptical disk by increasing $\Pe$. (e): similar to (d) but for a circular disk with a transition from an isotropic solute distribution. }
\label{fig:phase_concen}
\end{figure} 

We demonstrate in figure \ref{fig:phase_concen}(a) and (d) the bimodal distribution of normalized concentration 
caused by the curvature variation of the disk surface in the base state ($\Pe = 0.3$). 
The maximum normalized concentration $\Tilde{c}_{max}$ 
is located at the left vertex of the minor axis corresponding to $\beta = 90^{\circ}$, where $\beta$ denotes the polar angle with respect to $\be_s$. 
Here, we consider only the left half of the disk ($\beta \in \lb 0, 180^{\circ} \rb $) for its symmetry.  
At $\Pe = 0.5$, the base state loses stability and solutes are 
advected toward the disk's rear ($\beta = 180^{\circ}$). Correspondingly, the position of $c_{max}$ migrates rearward from $\beta = 90^{\circ}$ to the stagnation point $\mathrm{P}$, as depicted in figure \ref{fig:phase_concen}(b). 
Considering that the slip velocity $\bu_s = \nabla_s c$
is directed from low to high concentration at the disk surface, the fluid is attracted from the front and rear of the elliptical disk to the stagnation position $\mathrm{P}$, 
forming a puller-type swimmer (see figure \ref{fig:phase_puller}a). As $\Pe$ increases to $1.5$, the enhanced convective transport of solutes causes the shift from a bimodal to a polarized concentration profile with the peak concentration located at $\beta = 180^{\circ}$ (see figure \ref{fig:phase_concen}c). Correspondingly, the slip flow induced by the concentration gradient is driven from the front to the rear of the disk (see figure \ref{fig:phase_concen}b), generating a prototypical neutral-type swimmer. We comment that as an elliptical disk swims steadily, a bimodal concentration profile at its surface yields a puller-type swimmer, while a polarized one leads to a neutral-type counterpart. 
Different from the elliptical disk, a circular disk has an isotropic base distribution of the solute ($\Pe = 0.3$), as shown in figure \ref{fig:phase_concen}(e) and (g). 
For an unstable base state at $\Pe = 1.5$, solutes are advected rearward along the disk surface, leading to the direct transition from an isotropic to a polarized concentration profile. Hence, the circular disk behaves as a neutral swimmer solely.

 \subsection{Chiral symmetry-breaking orbiting motion} 
\begin{figure}[H]
\centering
\includegraphics[scale=0.7]{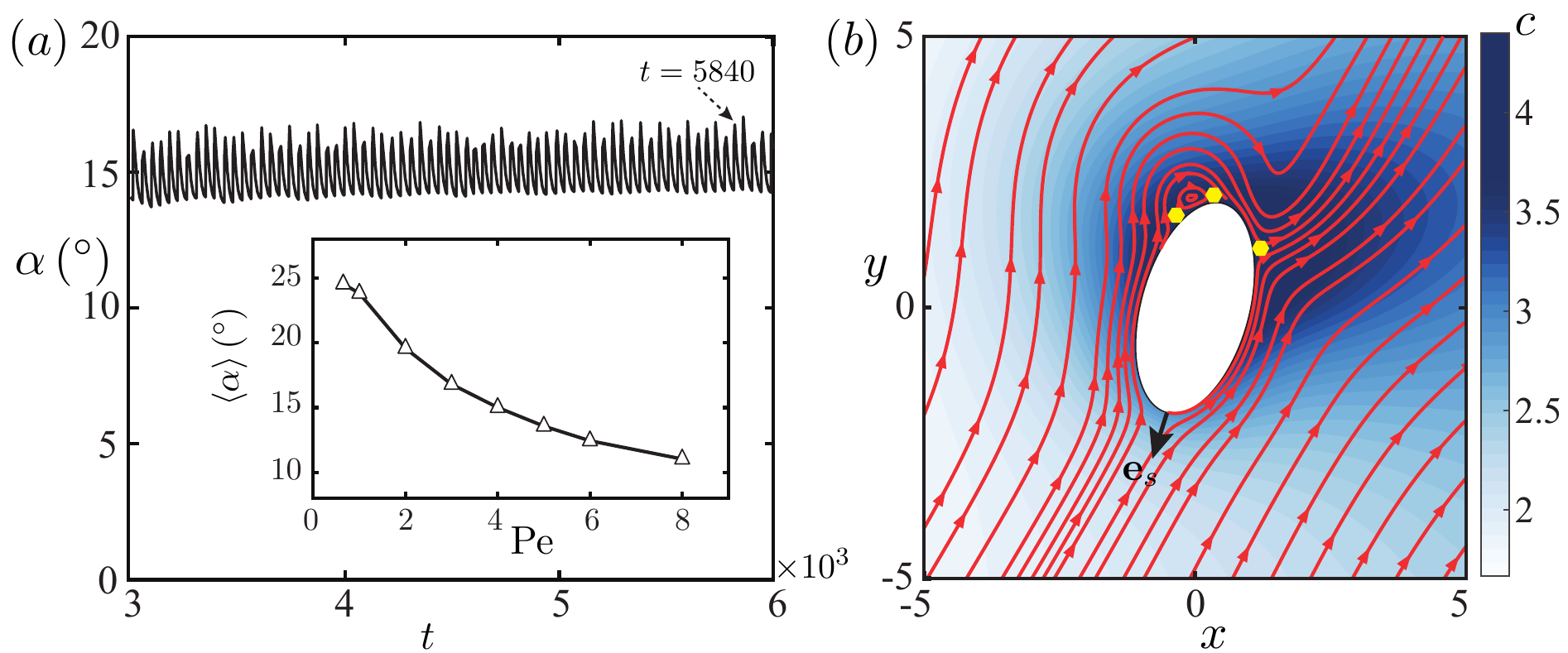}
\caption{(a) Time evolution of the angle $\alpha$ between the translational velocity $\bU$ and the disk orientation $\be_s$ as the elliptical disk with $e = 0.87$ approximately follows a circular trajectory for $\Pe = 4$. The inset shows the monotonic decrease of the time-averaged $\alpha$ with $\Pe$. 
(b) Solute distribution and streamlines at the instant $t = 5840$ marked in (a). Three hexagons denote the stagnation points at the rear of the disk.  
}
\label{fig:rotation}
\end{figure}

Having observed the puller-neutral transition in the steady swimming regime, we direct our focus on the chiral symmetry-breaking orbiting motion of the elliptical disk. Here, the disk approximately swims along a circular trajectory (see figure \ref{fig:trajec}b) with a rotational velocity $\Omega$. The time evolution of $\Omega$ resembles that of $\alpha$ in figure \ref{fig:rotation}(a) oscillating around a constant value: the constant value determines a globally circular trajectory and the oscillations contribute to a locally undulatory trajectory.
We find that the translational velocity $\bU$ (green arrow) deviates from $\be_s$ (red arrow) by an angle $\alpha$ oscillating slightly around $15^{\circ}$ for $\Pe=4$. 
In fact, the misalignment $\bU$ and $\be_s$ can be rationalized by examining the solute distribution $c(x,y)$ around the disk. Figure \ref{fig:rotation}(b) illustrates $c(x,y)$ and streamlines at a specific moment $t = 5840$.
The left-right asymmetry of the solute distribution about $\be_s$ gives rise to the asymmetric slip velocity, as reflected by the streamlines, hence causing diverging directions of $\bU$ and $\be_s$. 
Besides, we notice that three stagnation points near the rear of the disk in figure \ref{fig:rotation}(b) gradually migrate toward the rear with increasing $\Pe$ in the orbiting regime (not shown here). Accordingly, the time-averaged $\lt \alpha \rt = \int_0^{T} \alpha \mathrm{d}t /T$ within a time window $T$ decreases monotonically with $\Pe$, as shown in the inset of figure \ref{fig:rotation}(a). Intuitively, we infer $\bU$ is aligned with $\be_s$ ($\alpha = 0$) as three stagnation points exactly coincide at the rear of the disk, and indeed they do as $\Pe$ grows beyond $14$ where the disk executes a steady motion, e,g., $\Pe = 25$ (streamlines and concentration distribution are analogous to figure \ref{fig:phase_puller}b). 

The orbiting regime is characterized by the continuous 
rotation of an elliptical disk, which loses its chiral symmetry spontaneously. Conversely, a circular disk maintains this symmetry irrespective of the $\Pe$ value.
In particular, for $\Pe \in [9, 13]$, despite the asymmetric solute distribution and streamlines resembling those in figure \ref{fig:rotation}(b) occasionally, the circular disk only executes a meandering motion \citep{hu2019chaotic} without rotation. 
Ideally, a secondary stability analysis similar to the one performed above will help understand the mechanism for the shape-induced chiral symmetry breaking. This task is however technically challenging and will be pursued in the future.

Notably, experiments on active droplets in a Hele-Shaw cell 
showed that the droplet preferred to avoid the concentration trail emitted by itself at earlier times, termed a self-avoiding walk phenomenon \citep{hokmabad2021emergence}. 
The elliptical disk here does not avoid similarly 
but instead repeats its previous trajectory, as shown in figure \ref{fig:trajec}(b). The difference can be rationalized by a scaling analysis. We estimate that the time $\tb_d$ required for the concentration to decay to zero is of the order of $\bb^2/\Db$. The disk moves with the speed $\Umag \Vc$, hence the traveling time $\tb_p$ it takes to execute a circular trajectory of perimeter $\bar{L}_p$ is $ \bar{L}_p/\lp \Umag \Vc \rp$ approximately. 
Recalling the definitions of $\Vc$ and $\Pe$ in Section \ref{sec:prob}, we rewrite $\tb_p$ as $\bar{L}_p\bb/\lp \Umag \Pe \Db \rp$. The ratio between these two time scales is 
\begin{align}\label{eq:ratio_time}
\frac{\tb_p}{\tb_d} = \frac{\bar{L}_p/\bb}{\Umag\Pe}.
\end{align}
At a specific $\Pe$, e.g., $\Pe = 4$, in the orbiting regime, $\tb_p /\tb_d \approx  87.2 \gg 1$. This substantial ratio indicates that the disk swimmer can barely sense and thus avoid its own rapidly decaying chemical trace.  
In contrast, for a sufficiently larger $\Pe$, we may observe the self-avoidance due to the significantly reduced travel time $t_p$ \citep{hu2022spontaneous, hokmabad2021emergence}. 

 \subsection{Periodic swinging} 
\begin{figure}[H]
\centering
\includegraphics[scale=0.54]{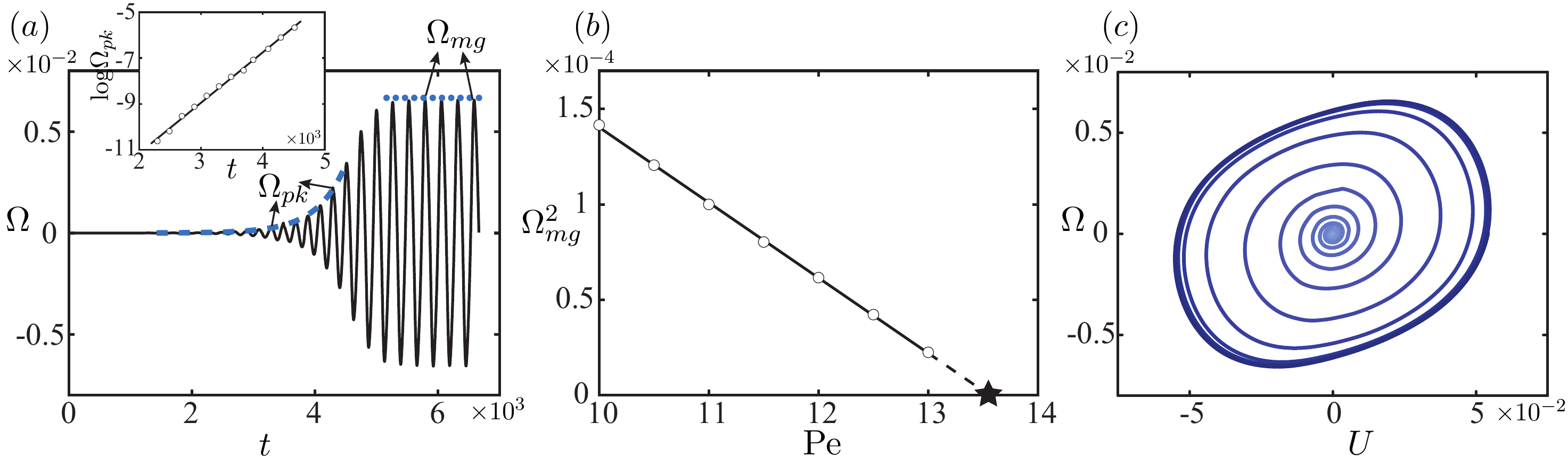}
\caption{(a) Time evolution of the rotational velocity $\Omega$ of an elliptical disk with $e = 0.87$ at $\Pe =12.5$. The dashed line denotes certain local peaks $\Omegpk$ of $\Omega$,
and the inset shows the linear dependence of $\mathrm{log}\;\Omegpk$ on $t$. (b) Linear variation of $\Omegmg^2$ in $\Pe$ near the critical $\Pe\approx 13.5$ (star), where $\Omegmg$ denotes the constant amplitude of $\Omega$ at $t > 5400 $, as depicted by the dotted line in (a). The periodic motion recovers to steady propulsion as $\Pe$ grows beyond the critical value. (c) Phase portrait in the $\Omega-U$ plane, with the color of the unstable spiral coded by $t$. }
\label{fig:hopf}
\end{figure}

The orbiting motion typically transitions to periodic swinging  (refer to figure \ref{fig:trajec}c) as $\Pe$ increases, which will be analyzed below.
Figure \ref{fig:hopf}(a) illustrates that the time-evolving rotational velocity $\Omega$ of the disk with $e = 0.87$ at $\Pe = 12.5$ is characterized by two phases. 
First, $\Omega$ grows rapidly due to self-oscillation, and the dashed line connecting local peaks $\Omegpk$ of $\Omega$ indicates an exponential growth of $\Omega$ in time. This trend is confirmed by the linear relationship between $\mathrm{log}\;\Omegpk$ and $t$ shown in the inset of figure \ref{fig:hopf}(a). Second, $\Omega$ saturates nonlinearly to a periodic state with a constant amplitude $\Omegmg$. The sinusoidal-like variation of $\Omega$ leads to a wave-like trajectory. We plot $\Omegmg^2$ as a function of $\Pe$ near the critical $\Pe \approx 13.5$ (star) in figure \ref{fig:hopf}(b). The linear dependence of $\Omegmg^2$ on $\Pe$ implies that the steady motion loses stability through a Hopf bifurcation. 
The critical $\Pe$ signifies the boundary between the periodic swinging and steady motion. We further present in figure \ref{fig:hopf}(c) the phase portrait in the $\Omega-U$ plane, where the unstable spiral at the origin grows continuously to an elliptical stable limit circle, indicating the supercritical nature of the Hopf bifurcation. 
Notably, when $e< 0.87$, we observe a transition from steady to periodic swinging with increasing $\Pe$, e.g., $e = 0.81$ (see figure \ref{fig:phase_diag}). The Hopf bifurcation still holds for the onset of instability that triggers this transition. 

\begin{figure}[H]
\centering
\includegraphics[scale=0.65]{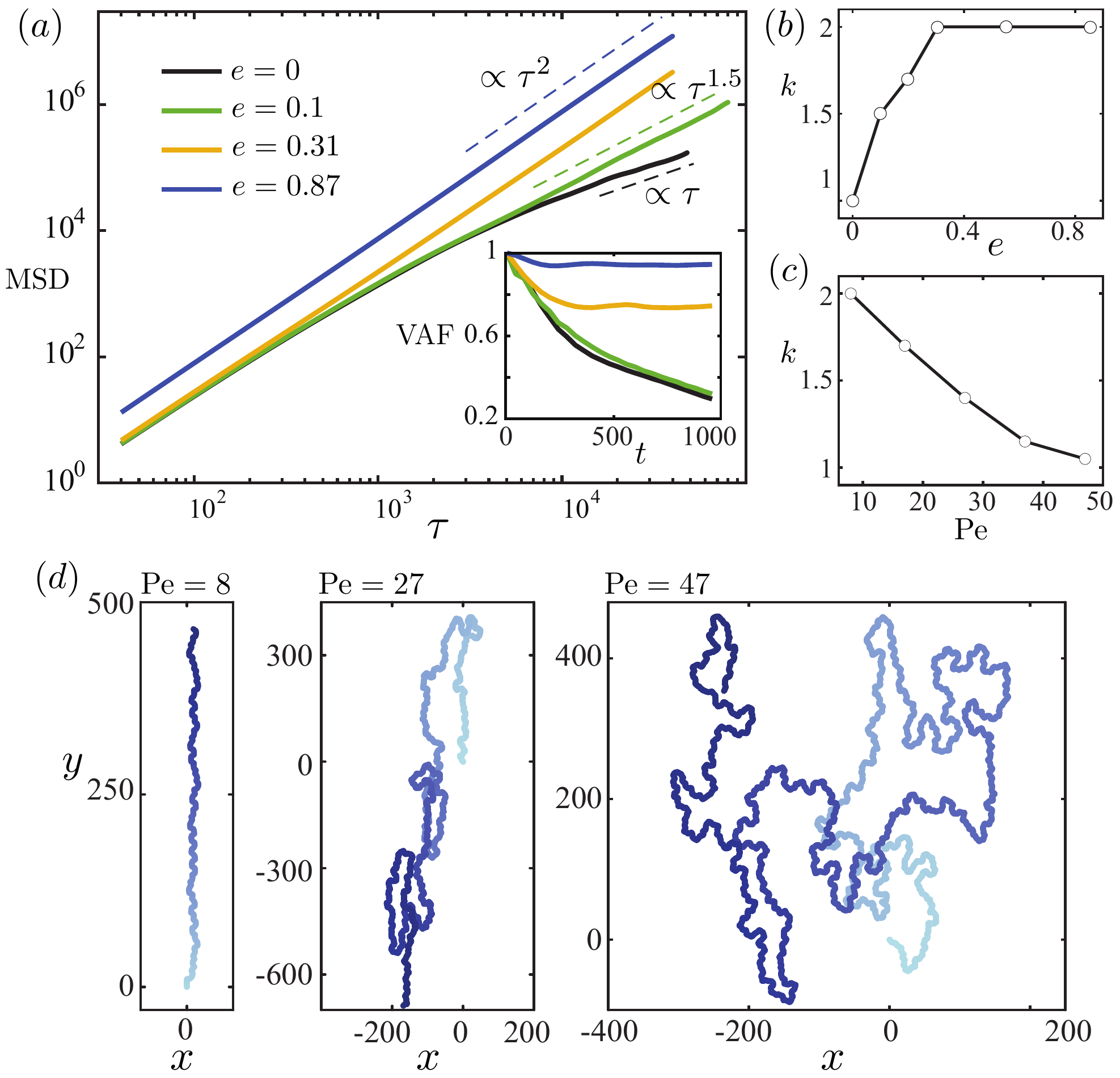}
\caption{(a) Mean square displacement (MSD) and velocity autocorrelation function (VAF) for disks with different shapes at $\Pectwo+10$, where $\Pectwo$ is the shape-dependent critical  P\'eclet number corresponding to the onset of chaos. (b) The exponent $k$ of the power-law scaling $\text{MSD} \propto \tau^k$ versus $e$. (c) Monotonic dependence of $k$ on $\Pe$ for the disk of $e = 0.2$. 
(d) Chaotic trajectories (color-coded by $t$) followed by an elliptical disk of $e = 0.2$ at $\Pe = 8$, $27$ and $47$, respectively.}
\label{fig:msd_cvv}
\end{figure} 

 \subsection{Chaotic swimming dynamics} 
Finally, we analyze the chaotic motion of elliptical disks by examining their mean square displacement (MSD) and the velocity autocorrelation function (VAF). We calculate the $\text{MSD} \lp \tau \rp$ of a disk 
\citep{michalet2010mean} that depends on a time lag $\tau$ by
\begin{equation}
    \text{MSD} \lp \tau \rp = \frac{1}{T-\tau}\int_{0}^{T-\tau}\lb \br\lp t+ \tau \rp -\br\lp t \rp \rb^2 \mathrm{d} t,
\label{MSD}
\end{equation}
where $\br \lp t \rp$ denotes the time-dependent displacement of a disk relative to its original position. Clearly, combining a large time period $T$ and $\tau \ll T$ can improve the statistical confidence. 
We calculate VAF using  
\begin{equation}
    \text{VAF} \lp t \rp = \frac{1}{T}\int_0^{T} \frac{\bU \lp t \rp \cdot\bU \lp t + \Delta t \rp}{ \ln \bU \lp t \rp \rn \ln \bU \lp t+ \Delta t \rp \rn}  \mathrm{d}t,
\label{Cvv}
\end{equation}
where a sufficiently large $T$ is used to attain statistical invariance \citep{chen2021instabilities}.  

We examine how the eccentricity $e$ of a disk affects its chaotic motion. First, we determine the shape-dependent $\Pectwo (e)$ when chaos emerges. Then, we investigate the chaotic dynamics at $\Pe(e) = \Pectwo (e) + 10$ above that threshold by a fixed offset, namely, $10$ here. The corresponding  MSD and VAF are depicted in figure \ref{fig:msd_cvv}(a). At early times, despite the diversity in the disk shape, they all swim persistently resulting in the quadratically growing  MSD in $\tau$. 
At long times, the disk shape considerably affects its phoretic movement, leading to a transition from a random-walking circular disk to a ballistically swimming elliptical counterpart. For the former, we reproduce perfectly its random walk behavior as reported by \citet{hu2019chaotic, lin2020direct, hu2022spontaneous}, which features the linear scaling $\text{MSD} \propto \tau$ and the decorrelation in velocity $\bU$ due to the rapidly changing swimming direction (see the inset). As the eccentricity $e$ increases to $0.1$, $\text{MSD} \varpropto \tau^{1.5}$, reminiscent of the self-avoidance walk identified experimentally~\citep{hokmabad2021emergence}.
We wonder whether this growth scaling, $\propto \tau^{1.5}$, results from the self-avoidance walk. 
By utilizing the scaling analysis mentioned above, we compare the decay time of the chemical $\tb_d$ and the traveling time $\tb_p$ of the disk and find that  $\tb_p /\tb_d \approx  40.8 \gg 1$ (see Equation~\ref{eq:ratio_time}), precluding self-avoidance here. 

At $e = 0.87$, the MSD's quadratic growth over lag time implies that the elliptical disk approximately executes the ballistic motion, as confirmed by the trajectory shown in figure \ref{fig:trajec}(e). Accordingly, the VAF is almost constant in time, indicating a correlation in velocity $\bU$. It is worth noting that \citet{hu2022spontaneous} and \citet{morozov2019nonlinear} also found the quadratic variations of MSD for the chaotic swimming of a phoretic particle and an active droplet, respectively, under the axisymmetric assumption.

By probing further in figure \ref{fig:msd_cvv}(b) the exponent $k$ of power-law scaling $\text{MSD} \propto \tau^k$ at varying $e$, we notice that the normal diffusion ($k = 1$) transitions to superdiffusion ($k > 1$) as $e$ increases. This observation demonstrates that the shape of a disk can significantly affect its diffusion behavior.  
In addition, we explore the effect of $\Pe$ on the chaotic motion of an elliptical disk at long times, as depicted in figure \ref{fig:msd_cvv}(c). In contrast to the eccentricity $e$, $k$ monotonically decreases with $\Pe$, implying the transition from superdiffusion to normal diffusion. Accordingly, the elliptical disk undergoes a ballistic motion near $\Pectwo$ and swims randomly at a larger $\Pe$, e.g., $\Pe = 47$ (see figure \ref{fig:msd_cvv}d). An analogous dependence of $k$ on $\Pe$ for the chaotic motion of an axisymmetric spherical particle was also reported by \citet{kailasham2022dynamics}. Note that the system exhibits ballistic evolution ($\text{MSD} \propto \tau^2$) at short times regardless of $\Pe$.

\section{Conclusions and discussions} \label{sec:conclu}
In this work, we numerically and theoretically investigate the swimming dynamics of an elliptical disk uniformly releasing solutes in the creeping flow regime. The disk with an eccentricity $e<0.75$ mimics a circular counterpart: it shows stationary, steady, periodic, and chaotic behaviors dependent on $\Pe$. When $e>0.75$, the disk attains an orbiting motion via an instability spontaneously breaking the chiral symmetry. 

By performing a LSA, we theoretically predict the critical P\'eclet number $\Pecone$ above which a stationary disk becomes a steady swimmer spontaneously triggered by instability. Besides the LSA,  we calculate the perturbation growth rate via a method akin to energy budget analysis.
The results agree with the LSA predictions and two contributing components of the growth rate showcase the competing roles of advection and diffusion.

We observe that the transition from a puller-type to a neutral-type steady disk swimmer is induced by a stronger rearward advection of solute due to growing $\Pe$. Accordingly, a bimodal concentration profile corresponding to a puller-type swimmer becomes a polarized profile leading to a neutral-type swimmer. 
The orbiting disk repeatedly swims along a circular trajectory while simultaneously rotating.

Two distinct types of periodic motions are identified: swinging and straight periodic motions. The former develops from a steady motion through a supercritical Hopf bifurcation and features a wave-like trajectory.  
The latter is characterized by a rectilinear motion with an oscillating swimming speed. 
Finally, the effects of the disk shape and $\Pe$ on the chaotic motion are examined. We uncover a shift from normal diffusion to superdiffusion with growing eccentricity $e$: the former and latter correspond to a random-walking circular disk and a ballistically swimming elliptical counterpart, respectively. The influence of $\Pe$ on the disk's diffusion behavior stands in contrast to that of $e$. 

It is worth mentioning that the prolate Janus droplet, as experimentally reported \citep{meredith2022chemical},  propels along its major axis resembling the behavior of our elliptical disk in the steady swimming regime. Nevertheless, the swimming orientation of the Janus droplet is determined by the inherent asymmetry of surface activity along its major axis, rather than by the instability-induced symmetry breaking. Additionally, we would like to emphasize that both the elliptical camphor disk \citep{kitahata2013spontaneous} and the prolate composite droplet \citep{hokmabad2019topological} in experiments demonstrated self-propulsion along their minor axes, rather than the major axis we observe. 
The differences may result from the failure of 2D simulations to capture the complex 3D physicochemical hydrodynamics present in experiments.  
Specifically, a swimming camphor disk is driven by Marangoni flow at the air-liquid interface and the resulting 3D subsurface flow, while the prolate droplet self-propels within a Hele-Shaw cell. We are planning 3D studies to address these discrepancies.

\begin{figure}[H]
\centering
\includegraphics[scale=0.55]{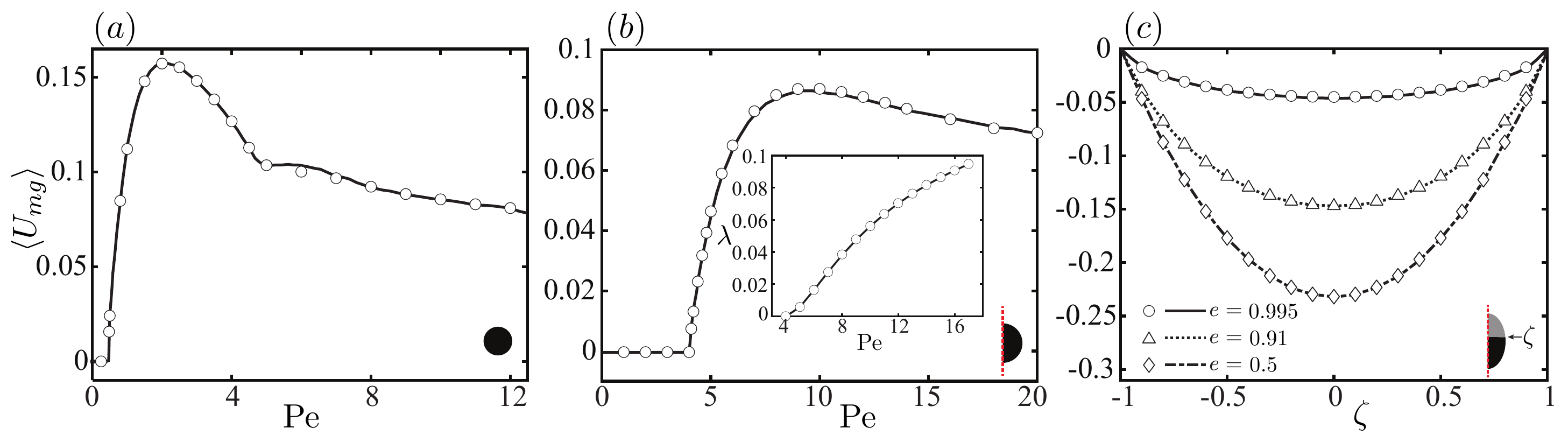}
\caption{Validation of our numerical implementation against published works. The numerical and published data are represented by markers and lines, respectively. 
(a) Swimming speed of an isotropic autophoretic disk in a circular domain of radius $R = 200$ for varying $\Pe$, benchmarked by ~\cite{hu2019chaotic}. Note that the time-averaged swimming speed $ \lt \Umag \rt = \int_0^{T} \Umag \;\nd t$ within a time period $T$ recovers $\Umag$ for steady propulsion. (b) Swimming speed of an autophoretic spherical particle versus $\Pe$ computed in an axisymmetric configuration, in comparison to~\citet{michelin2013spontaneous}; the inset shows the growth rate $\lambda$ of the unstable eigenmode versus $\Pe$.  (c) Swimming velocity of a spheroidal Janus particle as a function of $\zeta$, validated against \citet{popescu2010phoretic}. $\zeta$ denotes the height of the border dividing the particle into active and inert (marked in grey) compartments, with $\zeta = 0$ corresponding to the particle center. The eccentricity $e$ of the spheroid recovers to zero for a spherical particle.}
\label{fig:valida}
\end{figure} 

\appendix
\section{Validation of numerical implementations} \label{sec:appenA}
\begin{figure}[H]
\centering
\includegraphics[scale=0.55]{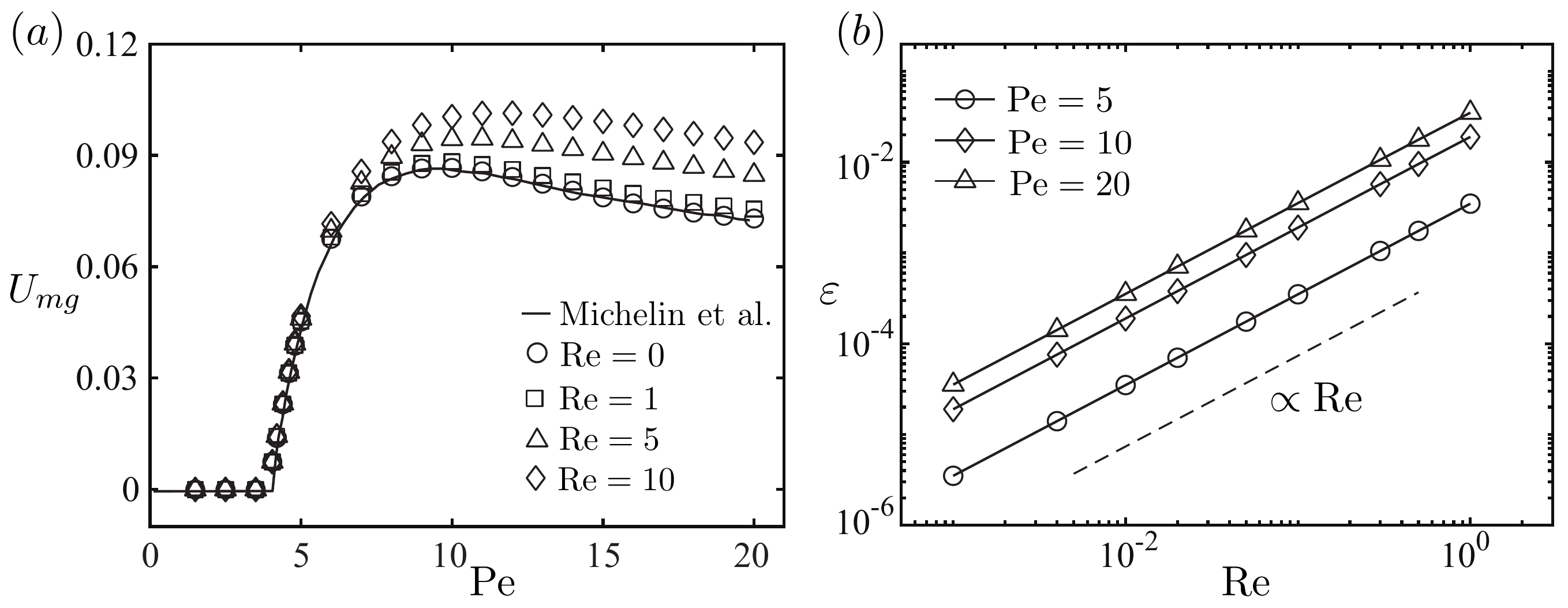}
\caption{(a) Swimming speed $\Umag$ of an isotropic phoretic particle versus $\Pe$ when $\Re$ varies, which is calculated in an axisymmetric setup. Our numerical data are compared with that of \citet{michelin2013spontaneous} at $\Re=0$. (b) Inertia enhances the swimming speed, characterized by the linear relation between the relative enhancement $\varepsilon= \left[ \Umag(\Re,\Pe) - \Umag (\Re=0,\Pe) \right] / \Umag (\Re=0,\Pe)$ and $\Re$. 
}
\label{fig:axis_sym_Re}
\end{figure}

We show the validation of our COMSOL implementations for the numerical simulation (Equations (\ref{NS})-(\ref{force_free})) and stability analysis (Equations \eqref{NS_h}-\eqref{force_torque_h}).
First, we study the inertialess self-propulsion of an isotropic autophoretic disk in a circular domain of radius $R = 200$. This setup has been investigated by  \citet{hu2019chaotic} combining a stability analysis and simulations based on a spectral method. The disk transits to a steady swimmer from its stationary state at the first threshold $\Pe \approx 0.466$. Increasing $\Pe$, the swimmer becomes unstable at the second critical condition $\Pe \approx 4.65$, moving in a meandering manner.
Our numerical data shown in figure \ref{fig:valida}(a) exactly recovers the reported $\Pe$-dependent swimming speed and two critical $\Pe$ values.
Besides the 2D case, we examine the spontaneous motion of an axisymmetric isotropic particle in an unbounded domain. In this case, instability occurs at $\Pe=4$ as predicted by \citet{michelin2013spontaneous}. Our data on swimming speed agrees well with the numerical results therein, as depicted in figure \ref{fig:valida}(b). Also, its inset shows that the present LSA recovers the threshold $\Pe=4$ and the eigenvalues depending on $\Pe$. As a side product, we probe the effect of inertia $\Re$ on this scenario. The inertial effect systematically enhances the swimming speed, becoming pronounced when $\Re > 1$ (see figure \ref{fig:axis_sym_Re}a). The inertia-induced relative enhancement of the swimming speed $\varepsilon$ in figure \ref{fig:axis_sym_Re}(b) is observed to linearly scale with $\Re$ in this weak inertia regime, which deserves further theoretical underpinning.

Finally, we examine the ability of our implementation in handling non-spherical phoretic swimmers. We focus on a spheroidal Janus colloid consisting of active and inert compartments. Its propulsion depends on its eccentricity $e$ and $\zeta$, where $\zeta$ denotes the height of the border separating the two compartments.
The obtained swimming speed agrees well with the reference solutions over a wide range of $e$ and $\zeta$ \citep{popescu2010phoretic}, as shown in figure \ref{fig:valida}(c).

\section{Periodic motion of an elliptical phoretic disk} \label{sec:appenB}

\begin{figure}
\centering
\includegraphics[scale=0.55]{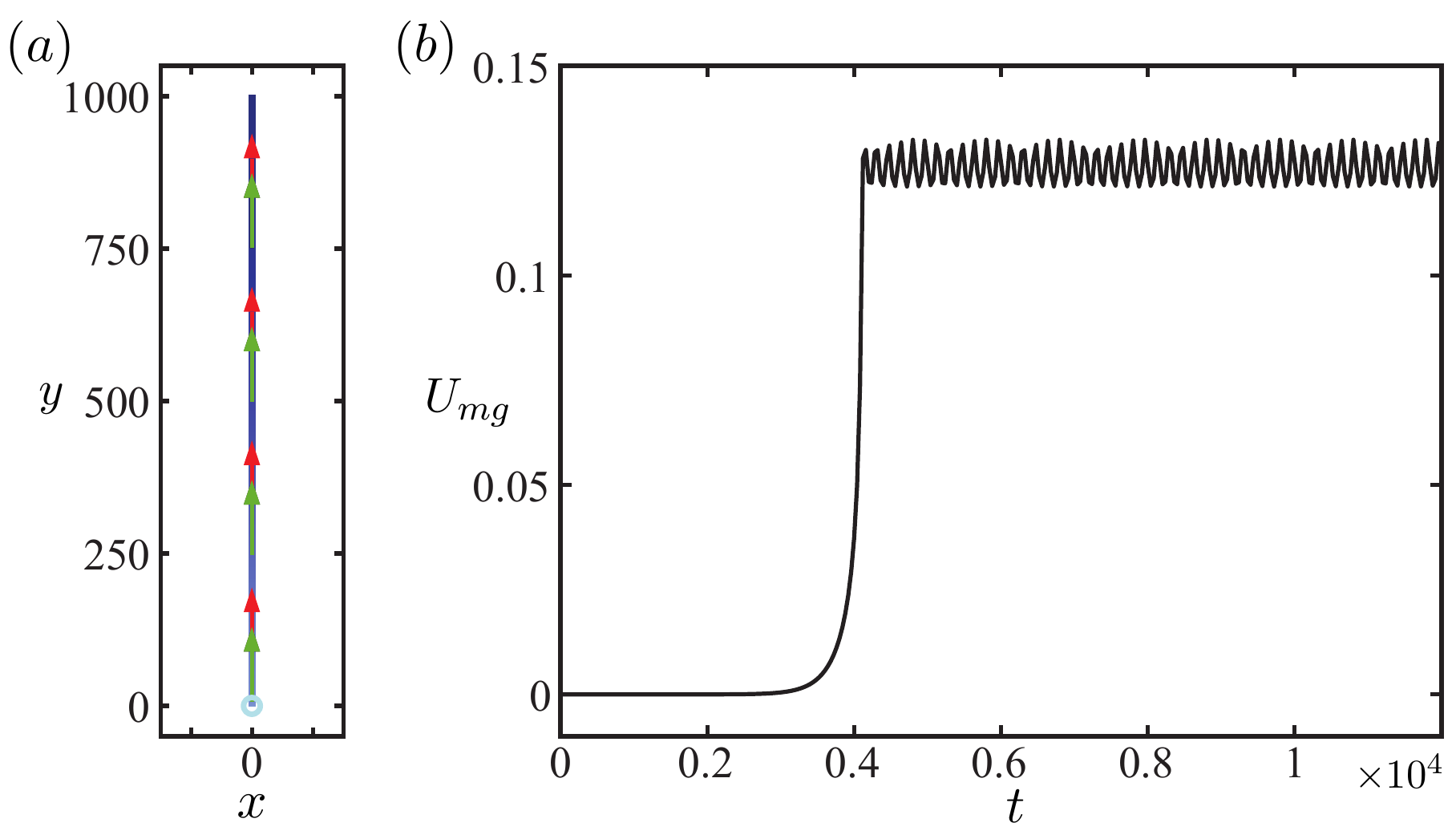}
\caption{Straight periodic motion of an elliptical disk with $e= 0.87$ at $\Pe =33$. (a) Straight trajectory color-coded by time $t$. 
The swimming direction $\be_s$ (red arrow) coincides with the translational velocity $\bU$ (green arrow). (b) Time evolution of the swimming speed $\Umag$. }
\label{fig:pe33}
\end{figure}

We identify two types of periodic motion: the first showcases a wave-like trajectory caused by the periodic rotation of the elliptical disk (see figure \ref{fig:trajec}g); the second corresponds to unidirectional rectilinear motion with a time-periodic swimming speed, as shown in figure \ref{fig:pe33}. 

\section{Dependency of swimming dynamics on the domain size $R$} \label{sec:appen}
\begin{figure}
\centering
\includegraphics[scale=0.55]{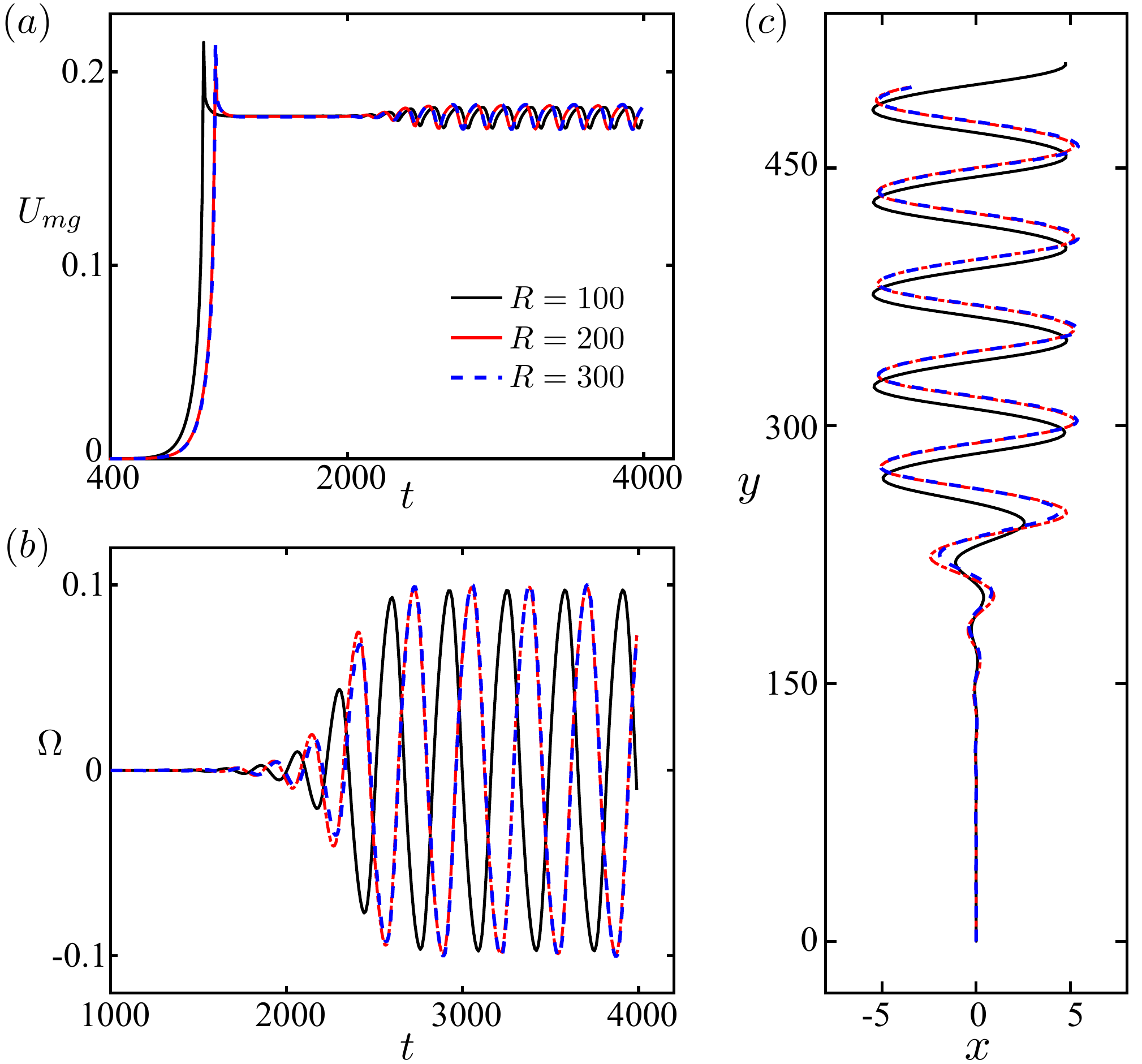}
\caption{(a) Swimming speeds, (b) rotational velocities, and (c) trajectories, of an elliptic disk with $e = 0.87$ and $\Pe = 11$ at varying $R$. Results at $R = 200$ and $R=300$ lie almost on top of each other. 
}
\label{fig:Rdep}
\end{figure}

In figure \ref{fig:Rdep}, we show the dependency of a disk's swimming dynamics on the domain size  $R$. The results obtained at $R=200$ agree perfectly with those at $R = 300$, suggesting that the selected size $R = 200$ should be sufficiently large.

\begin{acknowledgments}
We thank Qianhong Yang and Prabal Negi for helpful discussions, as well as Wei-Fan Hu for generously sharing the data utilized in the validation. We appreciate the insightful comments from the anonymous reviewers.  
L.Z. thanks Singapore Ministry of Education Academic Research Fund Tier 2 (MOE-T2EP50221-0012 \& MOE-T2EP50122-0015) and Tier 1 (A-8000197-01-00)  grants, and the Paris-NUS joint research grant (ANR-18-IDEX-0001 \& A-0009528-01-00). Some computation of the work was performed on resources of the National Supercomputing Centre, Singapore (https://www.nscc.sg).
\end{acknowledgments}


\end{document}